%% file: main.tex
\documentclass[journal]{IEEEtran}
\usepackage{amsmath,amsfonts}
\usepackage{algorithmic}
\usepackage{array}
\usepackage[caption=false,font=normalsize,labelfont=sf,textfont=sf]{subfig}
\usepackage{textcomp}
\usepackage{stfloats}
\usepackage{url}
\usepackage{verbatim}
\usepackage{graphicx}
\def\BibTeX{{\rm B\kern-.05em{\sc i\kern-.025em b}\kern-.08em
    T\kern-.1667em\lower.7ex\hbox{E}\kern-.125emX}}
\usepackage{balance}

\usepackage{overpic}
\usepackage{rotating}

\newtheorem{theorem}{Theorem}[section]
\newtheorem{proposition}{Proposition}[section]

\begin{document}
\title{ITS: Implicit Thin Shell for Polygonal Meshes}
\author{Huibiao Wen, Lei Wang, Yunxiao Zhang, Shuangmin Chen$^\dagger$, Shiqing Xin, \\Chongyang Deng, Ying He, Wenping Wang, Changhe Tu

\thanks{H. Wen, L. Wang, Y. Zhang, S. Xin, C. Tu are with School of Computer Science and Technology, Shandong University, Qingdao, Shandong, China. E-mail: ericvein@163.com, leiw1006@gmail.com, zhangyunxiaox@gmail.com, xinshiqing@sdu.edu.cn, chtu@sdu.edu.cn.}
\thanks{S. Chen is with School of Information and Technology, Qingdao University of Science and Technology, Qingdao, Shandong, China. E-mail: csmqq@163.com.}
\thanks{C. Deng is with School of Science, Hangzhou Dianzi University, Hangzhou, China. Email: dcy@hdu.edu.cn.}
\thanks{Y. He is with School of Computer Science and Engineering, Nanyang Technological University, Singapore. Email: yhe@ntu.edu.sg.}
\thanks{W. Wang is with Computer Science \& Engineering, Texas A\&M University, USA. E-mail: wenping@tamu.edu.}
\thanks{S. Chen is the corresponding author.}
}

\markboth{Journal of \LaTeX\ Class Files,~Vol.~18, No.~9, September~2020}%
{How to Use the IEEEtran \LaTeX \ Templates}
\maketitle
\begin{abstract}
In computer graphics, simplifying a polygonal mesh surface~$\mathcal{M}$ into a geometric proxy that maintains close conformity to~$\mathcal{M}$ is crucial, as it can significantly reduce computational demands in various applications. In this paper, we introduce the Implicit Thin Shell~(ITS), a concept designed to implicitly represent the sandwich-walled space surrounding~$\mathcal{M}$, defined as~$\{\textbf{x}\in\mathbb{R}^3|\epsilon_1\leq f(\textbf{x}) \leq \epsilon_2, \epsilon_1< 0,  \epsilon_2>0\}$. Here, $f$ is an approximation of the signed distance function~(SDF) of~$\mathcal{M}$, and we aim to minimize the thickness~$\epsilon_2-\epsilon_1$. To achieve a balance between mathematical simplicity and expressive capability  in~$f$, we employ a tri-variate tensor-product B-spline to represent~$f$. This representation is coupled with adaptive knot grids that adapt to the inherent shape variations of~$\mathcal{M}$, while restricting~$f$'s basis functions to the first degree. In this manner, the analytical form of~$f$ can be rapidly determined by solving a sparse linear system. Moreover, the process of identifying the extreme values of~$f$ among the infinitely many points on~$\mathcal{M}$ can be simplified to seeking extremes among a finite set of candidate points. By exhausting the candidate points, we find the extreme values~$\epsilon_1<0$ and $\epsilon_2>0$ that minimize the thickness. The constructed ITS is guaranteed to wrap~$\mathcal{M}$ rigorously, without any intersections between the bounding surfaces and~$\mathcal{M}$. ITS offers numerous potential applications thanks to its rigorousness, tightness, expressiveness, and computational efficiency. We demonstrate the efficacy of ITS in rapid inside-outside tests and in mesh simplification through the control of global error.
\end{abstract}


\input{body}
\end{document}

%% file: body.tex
\section{Introduction}\label{sec:Introduction}
\IEEEPARstart{P}{olygonal} mesh is a widespread form of 3D surface representation crucial in digital geometry processing. These meshes usually comprise a vast array of vertices, edges, and faces. Yet, direct manipulation or computational tasks on such meshes can be highly resource-demanding or sometimes rendered impractical due to issues with mesh quality. To overcome this challenge, a common approach involves employing a geometric proxy of reduced complexity, such as simplified meshes or boundary proxies, as noted in various studies~\cite{97GarlandSimplification, 15MandadSimplification, 19polygonalShell, 17boundaryProxy}.

Thin shells, defined as the enclosed regions between two bounding surfaces—often described as sandwich-walled spaces~\cite{23Shell}—provide notable benefits such as efficient memory usage~\cite{15Cage} and the ability to 
design curves on the meshes~\cite{23curvedesign}.
As a result, thin shells become common popular proxies and analysis tools for a wide range of applications, including 3D printing~\cite{17Printing}, solid texturing~\cite{07YeShellGeneration}, pattern engraving~\cite{23pattern-engraving},  surface and volume mappings~\cite{15Mapping}, meshing~\cite{15narrowbandcvt}, and cloth simulation~\cite{23Cloth}, among others.

The techniques for creating thin shells from a given polygonal mesh fall into two main categories: explicit and implicit methods. Explicit methods typically involve duplicating the base surface and then shifting it by a set distance. When it comes to choosing a direction for this displacement, there are several approaches, such as using normals~\cite{07YeShellGeneration, 04Shell, 05ShellMap} or the gradients of a generalized distance function~\cite{07HuangShell, 09Printing, 19CurveDesign, 20Shell}. Alpha wrapping~\cite{22Alpha}, a prominent explicit technique, differs by iteratively refining and carving a 3D Delaunay triangulation on an initially coarse enclosing surface of the input instead of merely offsetting triangles.

Implicit methods conceptualize the bounding surface of thin shells as a level set of a distance field, defining the thin-shell space with an implicit function and two levels of distance~\cite{02Shell,04ImplicitShell,11HanShell, 23MathMorphology}. Wang et al.~\cite{20WangContainmentCheck} proposed to represent the thin-shell of a polygonal surface for containment checks by using a collection of simple solids. Compared to explicit approaches, implicit methods provide more flexibility and uniformity, regardless of the quality of the input mesh. However, they often struggle to achieve a close and precise fit around the given mesh. Therefore, developing a tight, accurate implicit representation of thin shells continues to be a challenging area of research.

This paper focuses on addressing the challenge of representing sandwich-walled spaces using an implicit function, denoted as~$f$. While the signed distance function~(SDF) can indeed be interpreted as a tri-variable function, its lack of an analytic form poses challenges in computing the shell. Let~$\mathcal{M}$ be the input polygonal mesh. 
By approximating the SDF of~$\mathcal{M}$, we gain an analytic form that facilitates shell computation,
and the sandwich-walled space can be expressed as~$\{\boldsymbol{v}\in \mathcal{M}|f(\boldsymbol{v})\in[\epsilon_1, \epsilon_2]\}$, where $, \epsilon_1<0, \epsilon_2>0$ delineate the values characterizing the inner and outer implicit bounding surfaces respectively, rigorously enveloping the base surface~$\mathcal{M}$.
The desired implicit function must satisfy at least the following requirements:
(1)~\textit{Tightness}:~$\epsilon_1$ is as close to~$\epsilon_2$ as possible,
(2)~\textit{Rigorousness}:~$\epsilon_1\leq f(\mathcal{M})\leq \epsilon_2$,
(3)~\textit{Expressiveness}: even if the geometry/topology of~$\mathcal{M}$ is complicated, one can still find a suitable implicit function with a tight wrapping,
(4)~\textit{Rapid generation}: the generation of the implicit function~$f$ can be accomplished in a short time period, and the generation can scale well to handle large-scale data,
and
(5)~\textit{Rapid retrieval}: it can report the function value quickly with minimal computational effort.

In light of the requirements discussed, we present the Implicit Thin Shell (ITS) to represent the input's ambient space, which holds significant utility in~(1) meshless geometry processing, where the shell serves as the definition domain, and~(2) expedited proximity search, with the shell acting as the proxy.
ITS is designed to fulfill the outlined needs with a tri-variate tensor-product B-spline.
First, B-spline functions strike a balance between geometric expressiveness and computational efficiency. This characteristic is particularly beneficial for function evaluation, as it involves computations with only a relevant subset of parameters, thus speeding up the process. For the sake of simplicity and efficiency, we restrict these B-spline functions to the first degree.
Next, by adopting the established implicit B-spline methodology~\cite{15_3L}, we transform the signed distance function of a polygonal mesh surface into a tensor-product B-spline through a sparse linear system. 
Moreover, the task of identifying the extreme values of~$f$ with regard to the infinite points on~$\mathcal{M}$ is refined to a search for extremes among a finite set of candidate points, enabling rigorous wrapping. To further optimize efficiency and minimize memory usage, we employ the sparse voxel octree (SVO)~\cite{11SVO,10voxelization} instead of the uniform octree, without compromising on approximation quality.
We validate the utility of ITS in two key applications: conducting inside-outside tests and implementing mesh simplification with global precision control.

In summary, our contributions are three-fold:
\begin{itemize}
	\item We introduce a novel representation of the sandwich-walled space of an input surface, in which the implicit function is expressed as a tri-variate tensor-product B-spline.
	\item We reduce the problem of identifying the extreme values of $f$ among a countless number of points to seeking extremes among a finite set of candidate points. As a result, our ITS can rigorously wrap the input surface.
	\item We propose a set of acceleration strategies to enhance the implementation, 
 achieving significant improvements in runtime performance for inside-outside tests.
 We also utilize ITS as a tool to adapt the well-known QEM for mesh simplification, enabling global error control.
\end{itemize}

\section{Related Works}
In this section, we examine the prior research most relevant to our study, concentrating on two main areas: implicit surface reconstruction for thin-shell generation and the explicit creation of boundary cages.

\subsection{Implicit Representation}\label{Section:ISR}
Implicitization is a fundamental technique in digital geometry processing. For instance, the challenge of implicit surface reconstruction~\cite{06PSR} requires fitting a point cloud to a level surface defined by an implicit function. Given the extensive range of possible implicit functions, it is imperative to restrict the target function to a specific function space for it to be practically applicable. This restriction is commonly achieved by expressing the function as a linear combination of established basis functions~\cite{04MLS, 13Poisson}. Notable choices for these bases include B-spline bases~\cite{15_3L,17IHB,19Bspline}, radial basis functions (RBF)~\cite{01RBF,06RBF,10RBF}, wavelets~\cite{08wavelets,17wavelets}, and Fourier bases~\cite{05Fourier}. While there are several other methods for constructing implicit functions~\cite{18Reconstruct,20Reconstruction,22VoxelReconstruction,22ISROverview}, they are not detailed in this paper because their relevance to the main focus of this paper is comparatively minor.

\subsection{Explicit Generation of Boundary Cages}\label{Section:CAGE}
Bounding cages act as boundary proxies for polygonal meshes, serving as simplified external boundaries of thin shells. These cages have received considerable attention in cage-based deformation research~\cite{15Cage,18Deformation,DBLP:journals/tog/JuSW05}. Some techniques~\cite{08JuCage,13YangCage} generate cages by reusing a set of predefined models; however, these methods often struggle with generalization to new shapes. Voxelization of original meshes has proven effective in creating coarse cages~\cite{17boundaryProxy,09XianCage}. Oriented bounding boxes offer another strategy for cage construction~\cite{11XianCage}. Furthermore, some approaches~\cite{22Alpha,13StuartCage} start with a coarse outer boundary and iteratively refine it to tightly fit the original mesh. Alternatively, certain methods~\cite{09BenChenCage,11DengCage} focus on simplifying the dense offset surface to ensure the resulting cage closely wraps the original surface. While the explicit creation of boundary cages facilitates the production of outer boundary proxies with quality triangulation, the importance of inner bounding surfaces should not be underestimated in applications such as 3D printing~\cite{17Printing}, solid texturing~\cite{07YeShellGeneration,20Shell}
and cloth simulation~\cite{23Cloth}.

\section{Method}

\subsection{Problem Statement}
Let $\mathcal{M}$ represent a polygonal mesh surface. The first objective is to compute an implicit function~$f:\mathbb{R}^3\rightarrow\mathbb{R}$ to approximate the signed distance field of~$\mathcal{M}$: 
\begin{equation}\label{condition4surface}
	f(\boldsymbol{v})\approx 0, \qquad \forall \boldsymbol{v} \in\mathcal{M},
\end{equation}
such that~$f$ matches the signed distance function as closely as possible. It is preferable for the gradients of~$f$ being a unit vector, or at least they do not vanish. 
The second objective aims to precisely identify the extreme values of~$f$ for all points on~$\mathcal{M}$,  facilitating the delineation of the thin-shell space.

In Section~\ref{sec:Introduction}, we outline the five requirements for~$f$, 
namely tightness, rigorousness, expressiveness, rapid generation and rapid retrieval.
We quantitatively measure the tightness of~$f$ using the following formula:
\begin{gather}\label{initEnergy}
	\boldsymbol{E}(f) = \max_{\boldsymbol{v}\in\mathcal{M}} f(\boldsymbol{v})
	- \min_{\boldsymbol{v}\in\mathcal{M}} f(\boldsymbol{v}).
\end{gather}
Developing an efficient algorithm to compute~$f$ that meets all these requirements, given a polygonal surface as input, represents a challenging task.
The main challenge stems from the presence of infinitely many points on the mesh surface~$\mathcal{M}$, complicating the task of identifying the global extreme values of~$f$. To overcome this issue, we restrict~$f$ to a tri-variate tensor-product B-spline characterized by first-degree basis functions. The subsequent subsections will elaborate on the representation and construction of~$f$.

\subsection{Implicitization}
\label{subsec:implicitization}
Owing to its advantageous features, we intend to take the tensor-product B-spline to explicitly represent~$f$ that implicitly and approximately encodes the SDF.
As a result, the space of interest can be partitioned into regularly structured grid cells, enabling each basis function to be defined locally.
To achieve the \textit{rapid generation} and \textit{rapid retrieval} requirements as mentioned in Section~\ref{sec:Introduction}, we constrain the basis functions to be first-degree. 

The standard univariate first-degree B-spline is defined as
\begin{equation}
	B(t)=\left\{
	\begin{aligned}
		&1+t,  & &t\in[-1,0]\\
		&1-t,  & &t\in(0,1]\\
		&0,    & &\text{otherwise}
	\end{aligned}
	\right.,
\end{equation}
and the standard {tri-variate} tensor-product B-spline basis for any point~$\boldsymbol{v}\in\mathbb{R}^3$ with the coordinate~$(x,y,z)^\top$, rooted at the origin, can be written as
\begin{equation}
	\boldsymbol{B}(\boldsymbol{v})=B(x)B(y)B(z).
\end{equation}
If the side length of a grid cell is $w$, the tensor-product B-spline basis centered at the grid point~$\boldsymbol{g}$ becomes
\begin{equation}\label{eq:basis4grid}
	\boldsymbol{B}_{\boldsymbol{g}}(\boldsymbol{v})=\boldsymbol{B}\left(\frac{\boldsymbol{v}-\boldsymbol{g}}{w}\right).
\end{equation}
To this end, all the functions spanned by the basis functions defined at grid points have a form:
\begin{gather}
	f(\boldsymbol{v})=\sum_{\boldsymbol{g}\in\mathcal{G}} \lambda_{\boldsymbol{g}} \boldsymbol{B}_{\boldsymbol{g}}(\boldsymbol{v}),
\end{gather}
where $\mathcal{G}$ comprises all grid points and $\lambda_{\boldsymbol{g}}$ is the coefficient to be determined.

\input{FigTex/FigSVO}
\input{FigTex/FigBSpline}

In what follows, we shall introduce the method to divide the space into grid cells.
It is important to note that the accuracy of the representation is closely related to the resolution of the grid cells. Higher resolution results in higher accuracy. However, the number of grid cells increases cubically with respect to the linear degradation of~$w$.
This motivates us to implement voxelization of the space surrounding the polygonal mesh surface to minimize the grid count when defining implicit functions, as opposed to defining the function on the whole space. While utilizing implicit functions on voxelized grids can alleviate memory consumption, it might not completely meet the requirement for~\textit{expressiveness} when handling complicated meshes, particularly in achieving continuous inner or outer bounding surfaces for thin-shells.

To compensate for the limited expressiveness inherent in first-degree B-Splines and voxelized grids, a sparse voxel octree~(SVO) emerges as a superior option.
For pre-defined height~$K$ of the SVO, we begin by voxelizing the polygonal mesh into a set of grids with width~$2^{-K}$ initially, then proceed with the bottom-up construction of the SVO, as depicted in Figure~\ref{fig:svo}. 
Readers are encouraged to refer to~\cite{11SVO,10voxelization} for more details.
Herein, although we continue to use~$\mathcal{G}$ to denote the set of grid points, it is worth noting that two points at the same position but with different depths are considered distinct.
We associate a basis function for each grid point as defined in Equation~\ref{eq:basis4grid}.
With SVO, the function~$f$ can be formed as
\begin{gather}
	\label{eq:implicitfunction}
	f(\boldsymbol{v})=\sum_{k=0}^K\sum_{\boldsymbol{g}\in\mathcal{G}^k} \lambda_{\boldsymbol{g}} \boldsymbol{B}\left(\frac{\boldsymbol{v}-\boldsymbol{g}}{w_k}\right),
\end{gather}
where $w_k,\mathcal{G}^k,\boldsymbol{B}$
have similar definitions but are defined for the~$k$-th depth.
Figure~\ref{fig:Bspline} visualizes the first-degree univariate B-spline basis functions at different depths. 

Obviously, the detailed form of~$f$ is determined once the coefficients~$\{\lambda_{\boldsymbol{g}}\}$ are found. 
The computation of~$\{\lambda_{\boldsymbol{g}}\}$ 
involves incorporating each grid point~$\boldsymbol{g}\in\bigcup_k\mathcal{G}^k$ 
into the above representation, resulting in a system of linear equations,
each formed as:
\begin{gather}\label{eq:system}
	f(\boldsymbol{g})=\boldsymbol{D}(\boldsymbol{g},\mathcal{M}),\quad \forall \boldsymbol{g}\in \mathcal{G}^k,\;k=0,1,\cdots,K,
\end{gather}
where the right-hand side $\boldsymbol{D}(\boldsymbol{g},\mathcal{M})$ measures 
the signed distance from~$\boldsymbol{g}$ to the mesh surface~$\mathcal{M}$, and can be computed 
by fast winding number method~\cite{18LibiglWinding}.

\input{FigTex/FigSDF}

Because of the local support property of B-spline basis and the ability of children basis functions to reconstruct the parent basis function, the equation system is sparse and symmetric, but not positively definite. 
Although this may result in the system of Equation~\ref{eq:system} being singular and solutions being non-unique, a feasible solution can be obtained using the least squares conjugate gradient solver from the Eigen library.
To this end, we accomplish the implicitization of the signed distance function;
Figure~\ref{fig:SDF} validates the feasibility of the computed SDF.

\subsection{Extremity Detection}
\label{subsec:extremepoints}

To extremize the implicit function~$f$,
we consider a triangle~$T$ with vertices~$\boldsymbol{v}_1\boldsymbol{v}_2\boldsymbol{v}_3$ which lies entirely inside a grid cell at Depth-0. Any point~$\boldsymbol{v}\in T$ can be uniquely expressed as~$\boldsymbol{v} = (\boldsymbol{v}_2 - \boldsymbol{v}_1)\alpha + (\boldsymbol{v}_3 - \boldsymbol{v}_1)\beta + \boldsymbol{v}_1$, where~$0 \leq\alpha,\beta\leq 1$ and~$\alpha + \beta\leq 1$.
Therefore, Equation~\ref{eq:implicitfunction} can be rewritten by changing its variables from $\boldsymbol{v}$ to~$\alpha, \beta$.
We denote the new bivariate function as~$H_T(\alpha, \beta)$,
and $\nabla_{(\alpha, \beta)} H_T=\boldsymbol{0}$ provides part of the candidate points.
Along the boundary of $T$, $H_T(\alpha, \beta)$ can be further reduced
to a univariate function, formed like $\hat{H}_T(t)=H_T(\alpha, \beta)$,
subject to $\alpha=0$, $\beta=0$, or $\alpha+\beta=1$.
Therefore, it is likely that $\hat{H}_T'(t)=0$ provides some candidate points.
Finally, the three vertices of $T$ may also serve as the candidate points.
To summarize, the candidate points~$\boldsymbol{C}_T$, within the triangle~$T$, for extremizing~$f$ can be achieved by 
\begin{enumerate}
    \item $\nabla_{(\alpha, \beta)} H_T=\boldsymbol{0}$,
    \item $\hat{H}_T'(t)=0$, and
    \item the three vertices of $T$.
\end{enumerate}
More details are given in Appendix~\ref{sec:extreme}.

\begin{proposition}	\label{pro:containment}
	Given that 
	\begin{equation}	
		\mathcal{G}^0\subset\mathcal{G}^{1}\subset\cdots\subset\mathcal{G}^K,
	\end{equation}
	a grid cell at Depth-0 must be contained in a grid cell at the coarser level of SVO obviously.
\end{proposition}

Note that under the situation that triangle~$T$ intersects with grid lines or planes, $T$ can be split into sub-triangles such that each sub-triangle is entirely contained in certain grid cell at Depth-0 according to the Proposition~\ref{pro:containment}. 
Although these intersection points may also contribute to finding the extreme values,
they have been considered as the second or third candidate types, as mentioned above.

To this end, 
 the process of identifying the extreme values of~$f$ among the countless points on~$\mathcal{M}$ can be reduced to seeking extremes among~$\boldsymbol{C}$, which comprises a finite set of candidate points.
\begin{theorem}
	The thin-shell space defined by~$\epsilon_1\leq f(\mathcal{M})\leq\epsilon_2$ rigorously wraps~$\mathcal{M}$, where
	\begin{gather}
		\epsilon_1=\min_{\boldsymbol{v}\in\mathcal{C}}f(\boldsymbol{v}), \quad \epsilon_2=\max_{\boldsymbol{v}\in\mathcal{C}}f(\boldsymbol{v}).
	\end{gather}    
\end{theorem}

\textbf{Remark.}
It is well understood that locating a point within the sparse voxel octree (SVO) can be accelerated using Morton codes (i.e., the interleaved xyz coordinates). Additionally, B-splines are recognized for their advantageous property of local support. By leveraging these two features, the assessment of the values~$f$
can be achieved in with~$O(K)$, where~$K$ represents the depth of the SVO.

\input{FigTex/Figdetail}
\input{FigTex/FigThicknessTime} 

\input{FigTex/FigParameter}
\section{Experiments}

\subsection{Experimental Setting}
We implemented our algorithm using C++ on a desktop computer that is equipped with an Intel Core i9-13900K CPU running at 3.00GHz, 64 GB of RAM, and an NVIDIA GeForce RTX 4060 GPU. For solving the sparse linear system, we utilized the least squares conjugate gradient solver from the Eigen library~\cite{Eigen}. Additionally, to compute signed distances for the grid points of the sparse voxel octree, we employed the fast winding number method provided by Libigl~\cite{libigl}. To further accelerate the computation, we harnessed NVIDIA’s CUDA for constructing the SVO for ITS. 
For visualization, implicit surfaces were extracted as triangle meshes using the Marching Cubes algorithm~\cite{87mc}, with resolutions ranging from~$100^3$ to~$500^3$.
It is important to note, however, 
our output comprises a parameterized representation of the SDF, alongside an interval specifying the outer and inner shells. As a novel representation of the sandwich-walled space, there's no need for explicit extraction of the triangle mesh from the two bounding surfaces.

\subsection{Validation}
In our algorithm, the depth of the SVO directly influences both shell thickness and representation accuracy. We uniformly scale the input model to fit within a unit box and denote the depth of the SVO by a parameter~$K$, where~$K$ is a non-negative integer. Then we define the thickness of implicit thin shells as
	\begin{equation}
		\label{eq:thickness}
		d=\epsilon_1 - \epsilon_2.	
	\end{equation}

We begin by demonstrating the practical applicability of our algorithm through 500 random tests conducted on the Thingi10k dataset~\cite{Thingi10K}. We sampling~10K points on each mesh surface using blue noise sampling method~\cite{15JiangBluenoise}, and compute the corresponding approximated SDF values for $K=4, 5, 6,  7,  8, 9$. If the value of a point falls within the interval~$[\epsilon_1, \epsilon_2]$, we consider the point to be within the thin-shell. Specifically, for an open mesh, we regard the corresponding point as within the thin-shell when the value falls within the interval~$[0, \max\{|\epsilon_1|, |\epsilon_2|\}]$. The results of all the tests, expressed as the ratio of the number of points inside the thin-shell to the total sampling point number, are 100\%, demonstrating the validation of our algorithm.

Our approach utilizes sparse voxel grids, significantly improving time performance and enabling the processing of complex shapes efficiently. 
Figure~\ref{fig:detail} illustrates the precision with which the resulting thin shell accurately envelop the input model while mirroring the detailed features in the input model.
Figure~\ref{fig:thicknessTime} depicts the correlation between shell thickness and the parameter~$K$, as well as the associated time cost (measured in seconds) relative to~$K$.
We utilize the Sapphos-Head model from the Thingi10k dataset to showcase the thin-shell outcomes in relation to the height of the sparse voxel octree~(SVO), as illustrated in Figure~\ref{fig:parameter}. By incorporating candidate points into the function~$f$ and visualizing its values through slice-cut techniques, we find that a relatively large~$K$ enables our ITS to effectively mimic the original mesh, highlighting the potent expressive capacity of ITS. Furthermore, as~$K$ is increased, the produced thin-shell consistently captures intricate surface details, confirming its efficacy in preserving high-fidelity geometric features.

\subsection{Performance}
\input{FigTex/FigCAD}
\input{FigTex/FigDisconnected}
\input{FigTex/FigGenus}
\input{FigTex/FigDefect}
\input{FigTex/FigSoup}
\input{FigTex/FigCompare}

Our algorithm effectively processes CAD models with open boundaries and gaps, as demonstrated in Figure~\ref{fig:CAD}, where different patches are color-coded for clarity in thumbnails. 
Additionally, due to the inherent advantages of implicit representations, our algorithm accommodates disconnected components and models with high genus, as illustrated in Figure~\ref{fig:disconnected} and Figure~\ref{fig:genus}. Figure~\ref{fig:defect} showcases models exhibiting various defects, such as self-intersections and non-manifold edges. 
The configuration of the inner bounding surfaces effectively illustrates a constrained design, ensuring non-intersection with the input model.

For models that cannot be oriented, our approach offers the flexibility to approximate an unsigned distance field~(UDF) instead of a signed distance field~(SDF). Figure~\ref{fig:soup} displays the results when a triangle soup is used as input. This demonstrates the significant flexibility 
in supporting unoriented meshing objects. 
Nevertheless, this may lead to an undesired increase in the thickness of the thin-shell.

\subsection{Comparisons}

Alpha wrapping (AW)~\cite{22Alpha} refines and sculpts a 3D Delaunay triangulation on an offset surface of the input model in a greedy fashion. This method ensures the creation of a valid mesh that completely encloses the original mesh, even when dealing with inputs plagued by defects. However, AW is limited to computing only the outer layer and is susceptible to issues caused by broken holes. In contrast, our algorithm is capable of generating both inner and outer layers and adeptly fills open holes. For polygonal meshes with open holes (as shown in Figure~\ref{fig:compare}), treating the input as a closed mesh with defects allows us to approximate the SDF of the model effectively, thereby filling open holes. While AW can also fill holes, it tends to produce a coarser mesh that may not retain the model’s fine details. Conversely, treating the input as an open mesh, we approximate the UDF to achieve results similar to those of AW.
This demonstrates that our algorithm has the potential to accommodate the inherent defects.

While Jiang's method~\cite{20Shell} successfully generates both outer and inner bounding surfaces, it faces difficulties with meshes containing self-intersections or non-manifold edges/vertices. The containment check approach introduced in~\cite{20WangContainmentCheck} leverages the implicit representation of a simple solid to define the thin-shell of a polygonal surface, facilitating the assessment of whether triangles are inside or outside. Despite its applications in collision detection, the explicit extraction of its shell presents significant challenges.

\section{Application}
\input{FigTex/FigQuery}
\subsection{Inside-Outside Query}
Unlike traditional explicit thin shells, ITS employs a tri-variate tensor-product B-spline defined on a sparse voxel octree to represent the thin shell, facilitating immediate signed distance queries. In this work, ITS is utilized for conducting inside-outside tests, yielding highly accurate results except in cases where the function value is within the range of $[\epsilon_1,\epsilon_2]$. For such borderline scenarios, the query point may be classified as ``OnSurface'', or the Fast Closest Point Warping~(FCPW) technique~\cite{21FCPW} can be applied for a precise determination of the query outcome. 
FCPW, an efficient, header-only C++ library, utilizes a broad Bounding Volume Hierarchy~(BVH) with vectorized traversal to quicken the computation of nearest points and ray intersection queries, seamlessly integrating with ITS to improve the speed and precision of inside-outside queries.
SdfLib~\cite{23SDF} is a library meticulously designed to accelerate queries concerning signed distance fields derived from triangle meshes. This is achieved through the discretization of space, ensuring efficient processing of distance-related calculations.

Figure~\ref{fig:inside-outside-test} demonstrates our method's performance by randomly sampling~100K points within a box, whose dimensions progressively increase from 1$\times$ to 10$\times$ the size of the bounding box, all while maintaining a~$K$ value of~6. Our approach consistently delivers accurate inside-outside query results, assuming the accuracy of FCPW, except when query points are in close proximity to the mesh surface—though such instances are exceedingly rare. 

In Figure~\ref{fig:inside-outside-time}, we benchmark FCPW against two strategies utilizing a~$K$ setting of~6 for our method and a SIMD width of~4 for FCPW. Through comprehensive inside-outside testing on the Sapphos-Head model, we demonstrate that both strategies markedly surpass FCPW in speed. 
Meanwhile, we also compared with the maximum depth of octree of~6 for SDFLib.
The fast winding number method~\cite{18LibiglWinding} presents an alternative technique for computing the exact SDF of a triangle mesh or soup, utilizing the concept of winding number. However, this approach is associated with a considerable time overhead, as evidenced by our tests which indicated an average time of 53.55~$\mu s$, irrespective of the size of the sampling box.

\subsection{Mesh Simplification}
\input{FigTex/FigSimplification}

QEM~\cite{97GarlandSimplification} is a well-regarded technique for simplifying polygonal meshes through the iterative collapsing of edges. It utilizes a heap that prioritizes edge removal based on quadratic error metrics, celebrated for its ability to maintain mesh accuracy. However, QEM selects collapse candidates based solely on a greedy approach, often overlooking the cumulative impact on global error. To address this limitation, we have augmented QEM with the integration of ITS. Specifically, we use the thin shell as a criterion to assess whether newly generated vertices fall within the designated thin shell boundaries. If a new vertex is found to be outside the thin shell, the edge collapse operation is skipped, thereby ensuring a more globally consistent error management in mesh simplification.

Taking~$f$, as well as~$\epsilon_1, \epsilon_2$, as input, the simplification algorithm is described as follow steps:
\begin{itemize}
\item[1.] For each valid edge~$e\in E$, compute the preferred contraction target position~$\boldsymbol{v}_e$ of~$e$;
\item[2.] If~$\epsilon_1 < f(\boldsymbol{v}_e) < \epsilon_2$, insert~$e$ into a heap, ensuring that the operation with the minimum cost (quadratic error) is positioned at the top. Otherwise, we skip this operation;
\item[3.] Iteratively perform edge collapse with the least cost from the heap, 
generating new edge-collapse candidates in the 1-ring neighborhood of the new vertex,
and push the new candidate operations into the heap until the heap becomes empty or the target number of faces is achieved.
\end{itemize}   

Unlike QEM, which determines the termination condition based on a target number of faces, our approach utilizes the thin shell to dictate the degree of simplification applied to the input mesh. The key distinction between QEM and our method lies in our use of a global error evaluator, $f(\boldsymbol{v}_e)$, to regulate the priority of edge collapses, for preventing edge-collapse operations that may produce distant vertices. As Figure~\ref{fig:simPara} shows, 
the restriction imposed by the thin shell~($K=5$) allows for the preservation of more features while minimizing the approximation errors. 

An alternative way to control the priority of edge collapse 
is to use the following priority:
\begin{equation}
    \text{QEM}(e)+\gamma f^2(\boldsymbol{v}_e),
\end{equation}
where $\text{QEM}(e)$ is the local quadratic error and $\gamma$ is used to tune the influence of global deviation, which is set to 1 by default.
The latter term~$\gamma f^2(\boldsymbol{v}_e)$ is called \textbf{global term}.
In practice, we use the target number of faces to define the termination while disregarding~$\epsilon_1$ and $\epsilon_2$.
As the QEM only takes into account local errors during the simplification process, it tends to overlook significant global deviations.
For instance, in Figure~\ref{fig:sim}, QEM fails to retain the symmetrical structures presented in the original input, whereas we succeed in doing so.

\section{Conclusion}

In this paper, we introduce the Implicit Thin Shell (ITS) as a novel representation for the ambient space surrounding an input polygonal mesh. ITS is represented by a tri-variate tensor-product B-spline, anchored on a sparse voxel octree. This structure grants ITS remarkable expressiveness and the ability to scale efficiently with complex shapes. By narrowing the search of extreme values from an infinite to a finite set of candidate points, we establish a framework for finding the thin shell,  ensuring all points on the mesh fall within the thin-shell domain. Furthermore, we outline various strategies to enhance computational speed.
The distinct advantages of ITS—namely its precision, thoroughness, expressiveness, and computational efficiency—open up new avenues for application. These include conducting inside-outside tests and innovatively adapting the well-established Quadric Error Metrics (QEM) technique for mesh simplification, thereby enabling the management of global errors more effectively.

\textbf{Limitations.} Our algorithm shows inefficiency with models characterized by extremely thin structures, predominantly curved sections, or models that include narrow gaps. Under these circumstances, it becomes necessary to enhance the resolution of the sparse voxel octree (SVO), which in turn, leads to a significant increase in computational expense. 

\textbf{Future works.} We will aim to enhance the efficiency of ITS construction by swiftly eliminating candidate points that clearly do not influence the extreme values of~$f$. This will be achieved through the implementation of numerical approximation techniques. Moreover, leveraging the advantageous properties of ITS, we intend to investigate additional applications for ITS.


{\appendices
\section{Candidate points detection using Ferrari's solution}\label{sec:extreme}
For any triangle~$T$ with vertices~$\boldsymbol{v}_{1}, \boldsymbol{v}_{2}, \boldsymbol{v}_{3}$ of the base mesh, whose coordinates are defined as~$(x_1, y_1, z_1)^\top, (x_2, y_2, z_2)^\top, (x_3, y_3, z_3)^\top$ respectively, we assume that~$T$ is contained in an entire grid cell. 
Then any point lying on the triangle can be expressed as
\begin{equation}
	\boldsymbol{v} = \alpha\boldsymbol{v}_{1} + \beta\boldsymbol{v}_{2} + (1-\alpha-\beta)\boldsymbol{v}_{3} \quad 0\leq\alpha, \beta \leq 1, \alpha + \beta\leq 1,
\end{equation}
and its function value is 
\begin{equation}
	f(\boldsymbol{v})=\sum_{k=0}^K\sum_{\boldsymbol{g}\in\mathcal{G}^k} \lambda_{\boldsymbol{g}}B\left(\frac{x-x^{\boldsymbol{g}}}{w}\right)
	B\left(\frac{y-y^{\boldsymbol{g}}}{w}\right)B\left(\frac{z-z^{\boldsymbol{g}}}{w}\right)
\end{equation}
where, $(x,y,z)^\top$ are the coordinates of~$\boldsymbol{v}$.

We denote the new bivariate function as~$H_T(\alpha, \beta)= f(\boldsymbol{v})$,
due to the first-degree B-spline basis. Therefore, its gradient with respect to~$\alpha, \beta$ is quadratic, written as 
\begin{equation}\label{Eq::g_alpha}
	\left\{
	\begin{aligned}
		\nabla_{\alpha} H_T &= a_1\alpha^2+a_2\beta^2+a_3\alpha\beta+a_4\alpha+a_5\beta+a_6\\
		\nabla_{\beta} H_T &= b_1\alpha^2+b_2\beta^2+b_3\alpha\beta+b_4\alpha+b_5\beta+ b_6.
	\end{aligned}\right.
\end{equation}
For any pair~$(\alpha, \beta)$, if it satisfies~$\nabla_{(\alpha, \beta)} H_T=\boldsymbol{0}$, then we have 
\begin{equation}\label{Eq:relation-alpha-beta}
	\alpha=\frac{(a_1b_2 -a_2b_1)\beta^2+(a_1b_5-a_5b_1)\beta+(a_1b_6-a_6b_1)}{(a_3b_1-a_1b_3)\beta+a_4b_1-a_1b_4}.
\end{equation}
Substituting~$\alpha$ in Eq.(\ref{Eq::g_alpha}), we get a quartic equation with respect to~$\beta$, written as 
\begin{equation}
	\gamma_1\beta^4 + \gamma_2\beta^3 + \gamma_3\beta^2 + \gamma_4\beta + \gamma_5 = 0 
\end{equation}

Remove the cubic term. Let $\beta=u-\frac{\gamma_2}{4\gamma_1}$, then we will get a new quartic equation with respect to $u$, $u^4+\omega u^2+\mu u+\nu = 0$, where 
$\omega = \frac{-3\gamma_2^2}{8\gamma_1^2} + \frac{\gamma_3}{\gamma_1}$, $\mu =  \frac{\gamma_2^3}{8\gamma_1^3}- \frac{\gamma_2\gamma_3}{2\gamma_1^2}+ \frac{\gamma_4}{\gamma_1}$, $\nu = \frac{-3\gamma_2^4}{256\gamma_1^4} + \frac{\gamma_2^2\gamma_3}{16\gamma_1^3} - \frac{\gamma_2\gamma_4}{4\gamma_1^2} + \frac{\gamma_5}{\gamma_1}$. In this way, if we want the right-hand formula to be a quadratic term, we have $\Delta = \mu^2 - 4(2y-\omega)(y^2-\nu)=0\Rightarrow y^3-\frac{\omega}{2}y^2-\nu y +\frac{\omega}{2}\nu - \frac{\mu^2}{8} = 0$. The solution $y$ of this equation can be gained easily with Newton's method, then we have $(u^2+y)^2 = (u - \frac{\mu}{4y-2\omega})^2$. Obviously, there is an easy way to get the result 
\begin{equation}
	\beta = \frac{\pm1\pm\sqrt{1-4(y \pm \frac{\mu}{4y-2\omega})}}{2} - \frac{\gamma_2}{4\gamma_1} 
\end{equation}
According to Eq.\ref{Eq:relation-alpha-beta}, if $\alpha, \beta\in [0, 1]$ and $\alpha + \beta \leq 1$, the point $\boldsymbol{v}$ will be the candidate point. 

Along the boundary of $T$, $H_T(\alpha, \beta)$ can be further reduced
to a univariate function, formed like $\hat{H}_T(t)=H_T(\alpha, \beta)$,
subject to $\alpha=0$, $\beta=0$, or $\alpha+\beta=1$.
Therefore, it is likely that $\hat{H}_T'(t)=0$ provides some candidate points.
Finally, the three vertices of $T$ may also serve as the candidate points.

\bibliographystyle{IEEEtran}
\bibliography{main.bib}



%% file: FigTex/FigSVO.tex
\begin{figure*}
\centering
	\graphicspath{{Figures/}}
	\begin{overpic}
		[width=18cm]{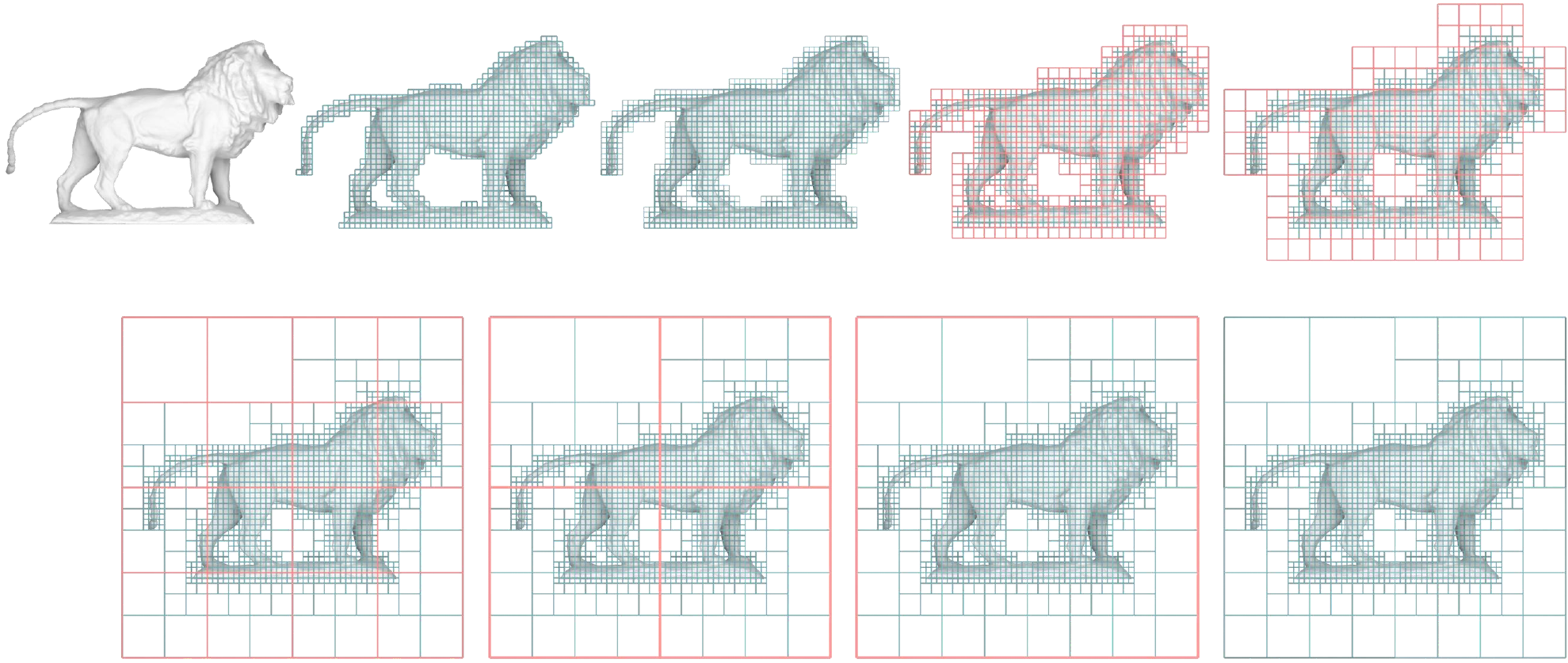}
		\put(7.0,23.4){(a)~Input}
		\put(22.45,23.4){(b)~Voxel grids}
		\put(41.3,23.4){(c)~Depth-0 grids}
		\put(61.0,23.4){(d)~Depth-1 grids}
		\put(82.0,23.4){(e)~Depth-2 grids}
		
		\put(2.0,10.9){\Huge{...}}
		\put(10.3,-1.9){(f)~Depth-(K-2) grids}
		\put(33.5,-1.9){(g)~Depth-(K-1) grids}
		\put(58.4,-1.9){(h)~Depth-K grids}
		\put(85.5,-1.9){(i)~SVO}
	\end{overpic}
	\vspace{0.3cm}
	\caption{Construction of sparse voxel octree~(SVO) in a bottom-up manner for a 2D case. Initially, an input polygonal mesh~(Figure~a) needs to be voxelized into a set of grids with width~$2^{-K}$~(Figure~b). Subsequently, the minimum number of grids with width~$2^{-K}$ must be added to the set~(Figure~c), ensuring that every four adjacent grids can form a larger grid with width~$2^{1-K}$~(the red grids in Figure~d).  This process continues from Depth-2 to Depth-K ~(from Figures~e to~h) with grids being constructed in the same manner, resulting in the hierarchical structure of the SVO~(Figure~i).}
	\label{fig:svo}
\end{figure*}

%% file: FigTex/FigBSpline.tex
\begin{figure}
	\centering
	\begin{overpic}
		[width=8.5cm]{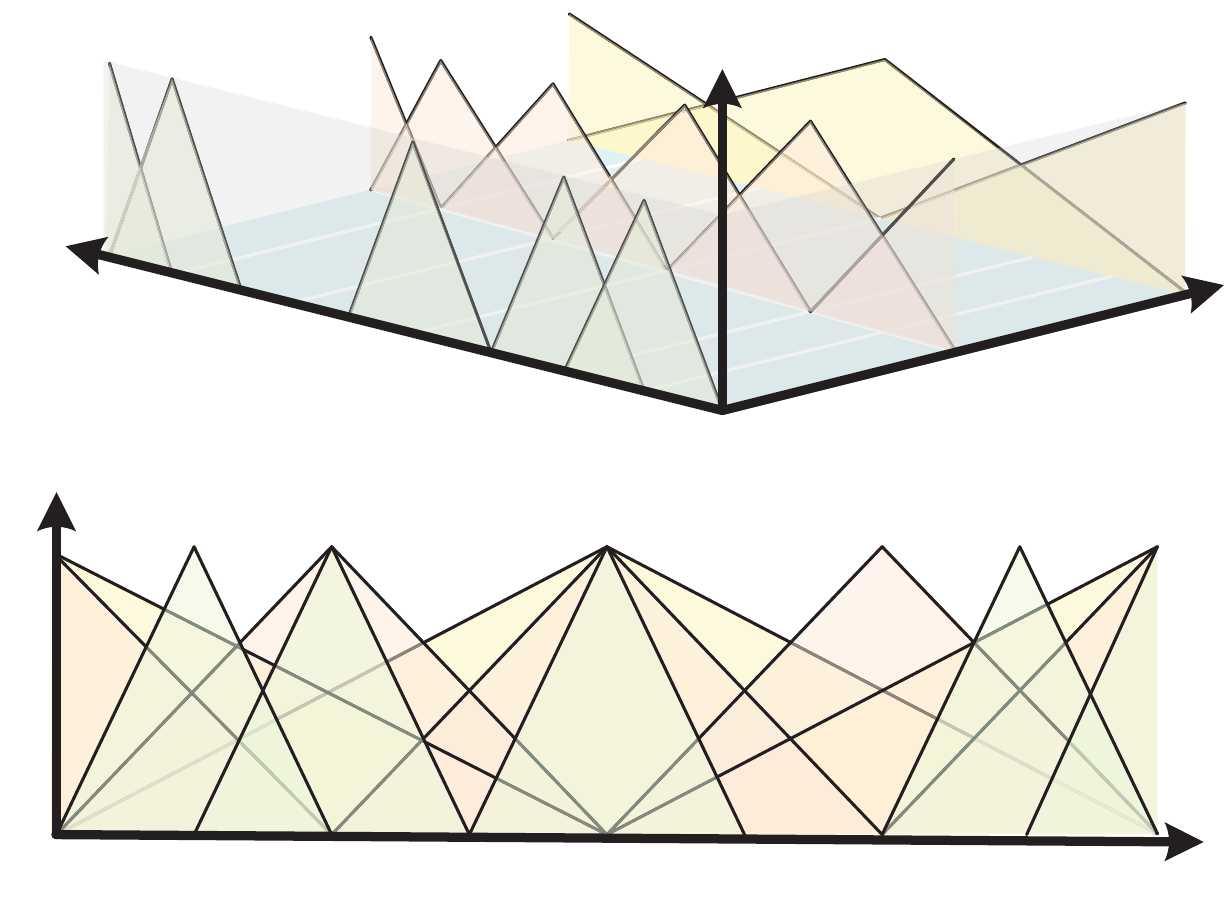}
		\put(7,50){\begin{turn}{-15}grid  points\end{turn}}
		\put(90,44){\begin{turn}{17}depth\end{turn}}
		\put(0,28){1}
		\put(82,1){grid points }
	\end{overpic}
	\caption{
		Top: First-degree univariate B-spline basis functions at different depths (visualized in different colors).
		Bottom: An overlay view of the basis functions at different depths. 
	}
	\label{fig:Bspline}
\end{figure}

%% file: FigTex/FigSDF.tex
\begin{figure}
\centering
	\graphicspath{{Figures/}}
	\includegraphics[width=8.5cm]{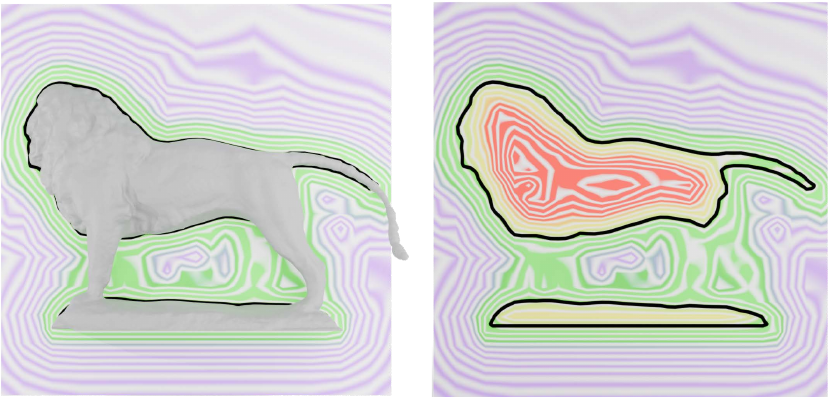}
	\caption{The approximate SDF of the polygonal mesh using B-Spline functions. The 0-level set curve colored in black closely aligns with the model along the cutting plane boundary, achieving a high degree of alignment with concordance.}
	\label{fig:SDF}
\end{figure}

%% file: FigTex/Figdetail.tex
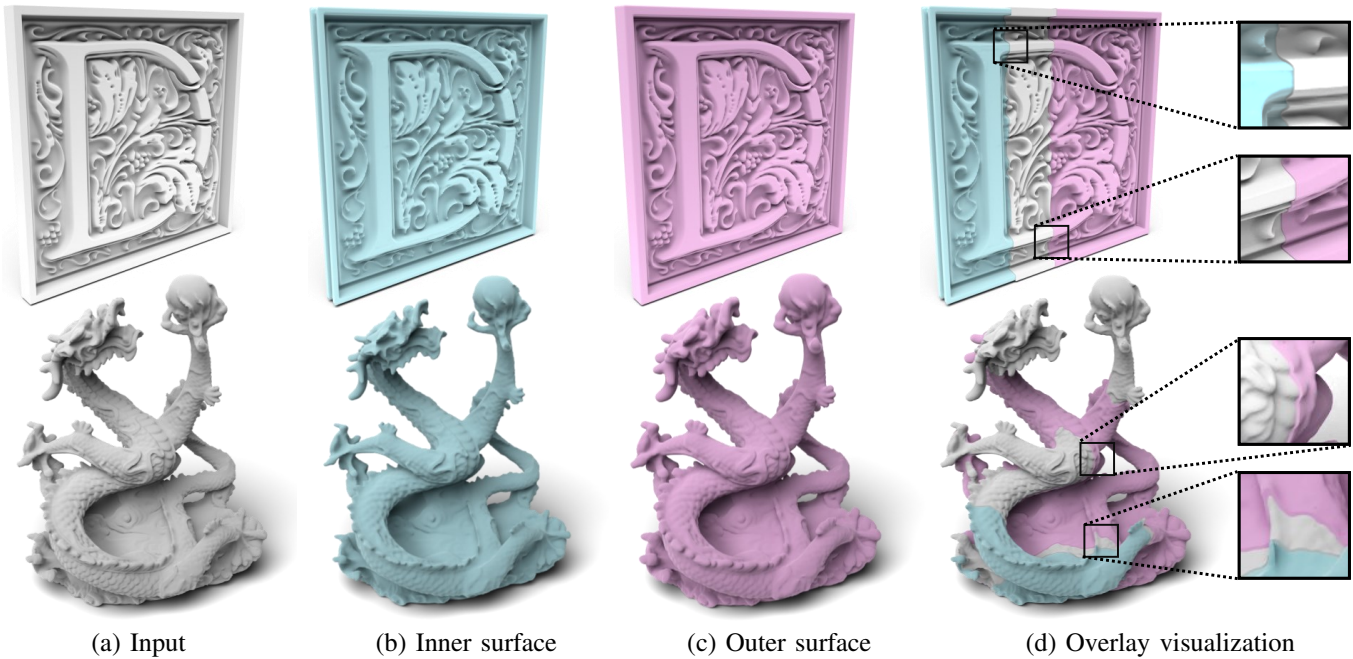
\begin{figure*}[t]
    \graphicspath{{Figures/details/}}
	\centering
	\begin{overpic}
		[width=18cm]{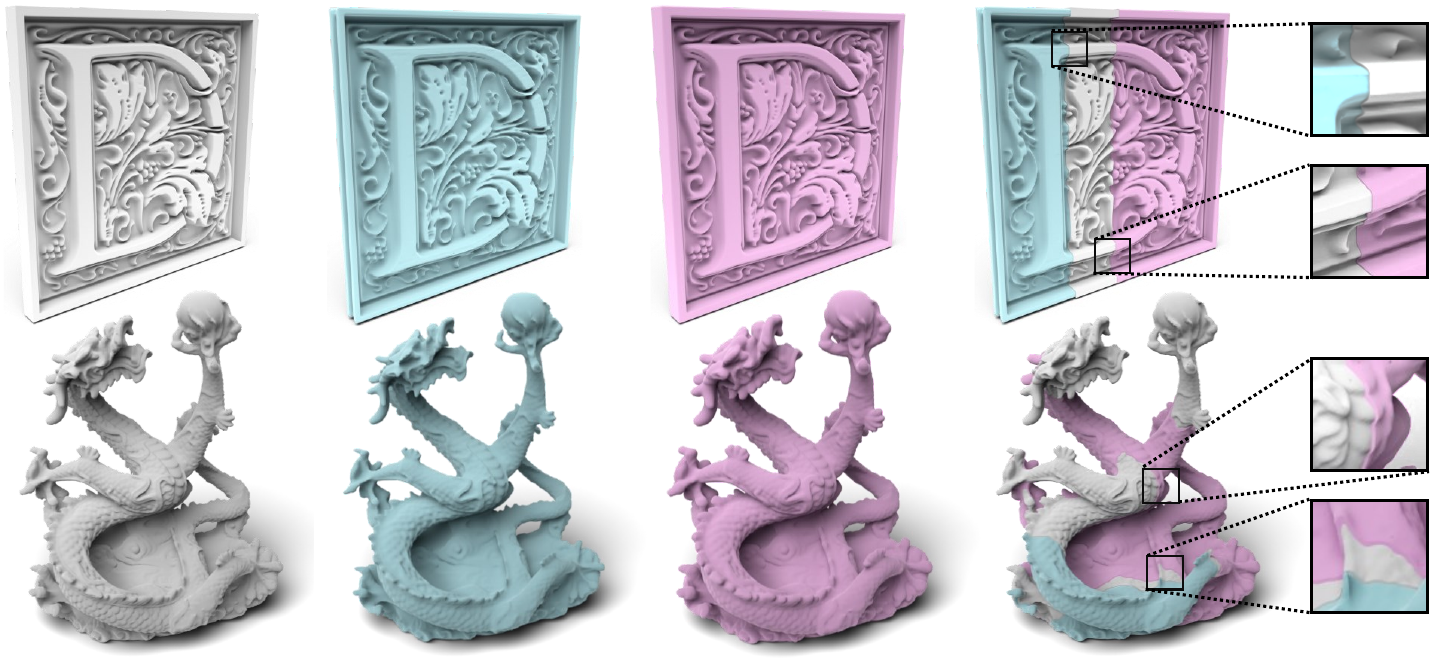}
	
		\put(6.8,-2.0){(a)~Input}
		\put(27.8,-2.0){(b)~Inner surface}
		\put(50.8,-2.0){(c)~Outer surface}
		\put(75.8,-2.0){(d)~Overlay visualization}
	\end{overpic}
	\vspace{0.25cm}
	\caption{
  ITS is particularly effective for models with abundant levels of detail. The inner and outer bounding surfaces of the thin shell showcase details and features that closely resemble those of the input model. In Figure~(d), it is evident that the model is enveloped with precision.
	}
	\label{fig:detail}
\end{figure*}

%% file: FigTex/FigThicknessTime.tex
\begin{figure}
	\graphicspath{{Figures/}}
	\centering
	\begin{overpic}
		[width=8.2cm]{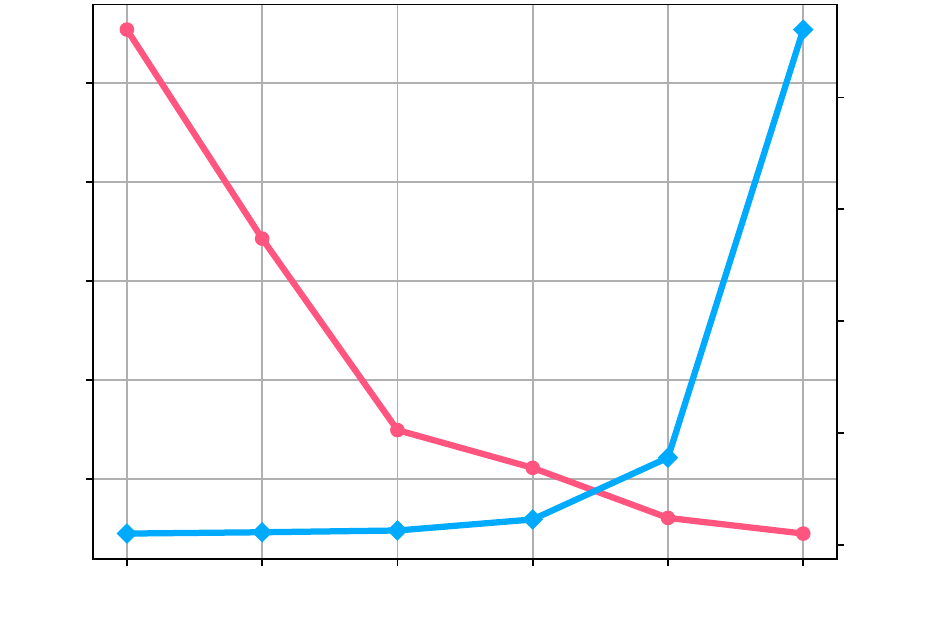}
		\put(-4.0, 30){\begin{turn}{90}Thickness \end{turn}}
		\put(98, 16){\begin{turn}{90}Execution time~(seconds) \end{turn}}

        \put(15.8, 64.0){\textbf{thickness}}
		\put(76.0, 64.0){\textbf{time}}
        
		\put(1.0, 58.3){0.05}
		\put(1.0, 47.6){0.04}
		\put(1.0, 36.9){0.03}
		\put(1.0, 26.5){0.02}
		\put(1.0, 15.6){0.01}

        \put(92.0, 56.8){60}
		\put(92.0, 44.9){45}
		\put(92.0, 32.9){30}
		\put(92.0, 20.9){15}
		\put(92.0, 9.0){0}

        \put(13.0, 2){4}
		\put(27.2, 2){5}
		\put(42.0, 2){6}
		\put(56.3, 2){7}
		\put(71.1, 2){8}
		\put(85.4, 2){9}
	\end{overpic}
	\vspace{-0.1in}\caption{
		Consider an input model normalized to fit within a unit box, comprising~32794 faces and~16323 vertices. The height of the SVOs is denoted by~$K$. We plot the dependence of thickness on the parameter~$K$, as well as the running time~(measured in seconds) with respect to~$K$.
	}
	\label{fig:thicknessTime}
\end{figure}

%% file: FigTex/FigParameter.tex
\begin{figure*}
	\graphicspath{{Figures/}}
	\centering
	\begin{overpic}
		[scale=0.6]{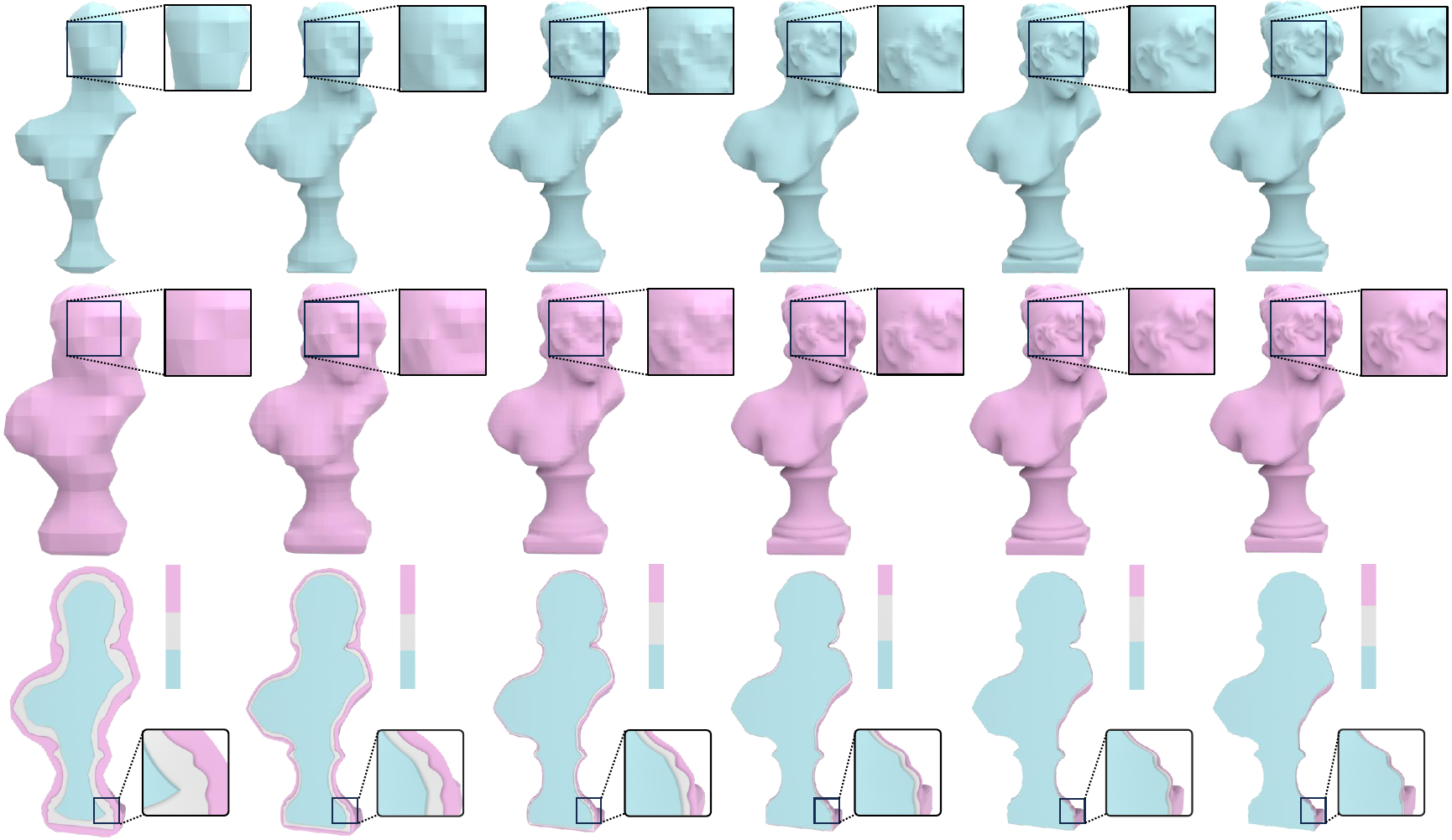}
		\put(12.6, 12.8){\small{-0.027}}
		\put(13.1, 15.3){\small{0}}
		\put(13.1, 18.5){\small{0.034}}
		
		\put(28.8, 12.8){\small{-0.013}}
		\put(29.3, 15.2){\small{0}}
		\put(29.3, 18.5){\small{0.017}}
 
 		\put(46.0, 13.2){\small{-0.008}}
 		\put(46.5, 15.9){\small{0}}
 		\put(46.5, 18.5){\small{0.007}}
 		
 		\put(61.7, 13.5){\small{-0.005}}
 		\put(62.2, 16.4){\small{0}}
 		\put(62.2, 18.5){\small{0.003}}
 		
 		\put(79.2, 13.4){\small{-0.003}}
 		\put(79.7, 16.4){\small{0}}
 		\put(79.7, 18.5){\small{0.002}}
 		
 		\put(95.1, 13.1){\small{-0.001}}
 		\put(95.6, 15.8){\small{0}}
 		\put(95.6, 18.5){\small{0.001}}
	\end{overpic}
	\caption{Given that the thin-shell structure has an inner layer~(top row) and an outer layer~(middle row), we visualized the results using the Sapphos-Head model as input. From left to right:~$K = 4, 5, 6, 7, 8, 9$. By incorporating candidate points into~$f$ , we visualized the function value using a slice-cut style, as shown in the bottom row. When~$K$ is relatively large, our ITS is expressive enough to closely approximate the original mesh. For example, when~$K=9$, the maximum absolute value of~$f$ is~$1\times 10^{-3}$, representing a negligible deviation from the original base surface.}
	\label{fig:parameter}
\end{figure*}

%% file: FigTex/FigCAD.tex
	\begin{figure}
		\centering
		\includegraphics[width=8.5cm]{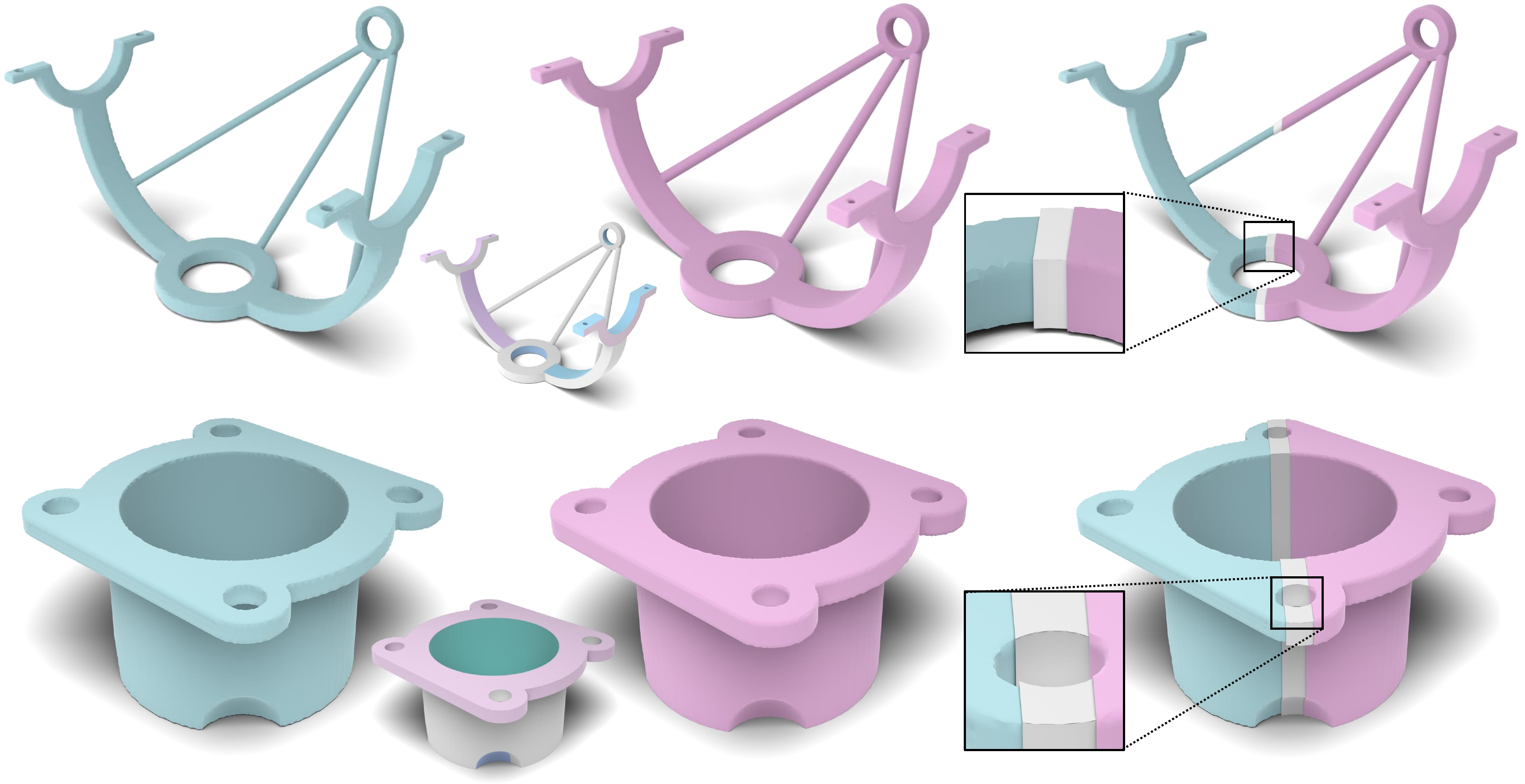}
		\vspace{-0.1in} \caption{
		ITS for surfaces with multiple patches (visualized in different colors; see the thumbnails). From left to right: the inner bounding surface, the outer surface, and the overlay visualization. 
		}
		\label{fig:CAD}
	\end{figure}

%% file: FigTex/FigDisconnected.tex
\begin{figure}
	\centering
	\includegraphics[width=8.5cm]{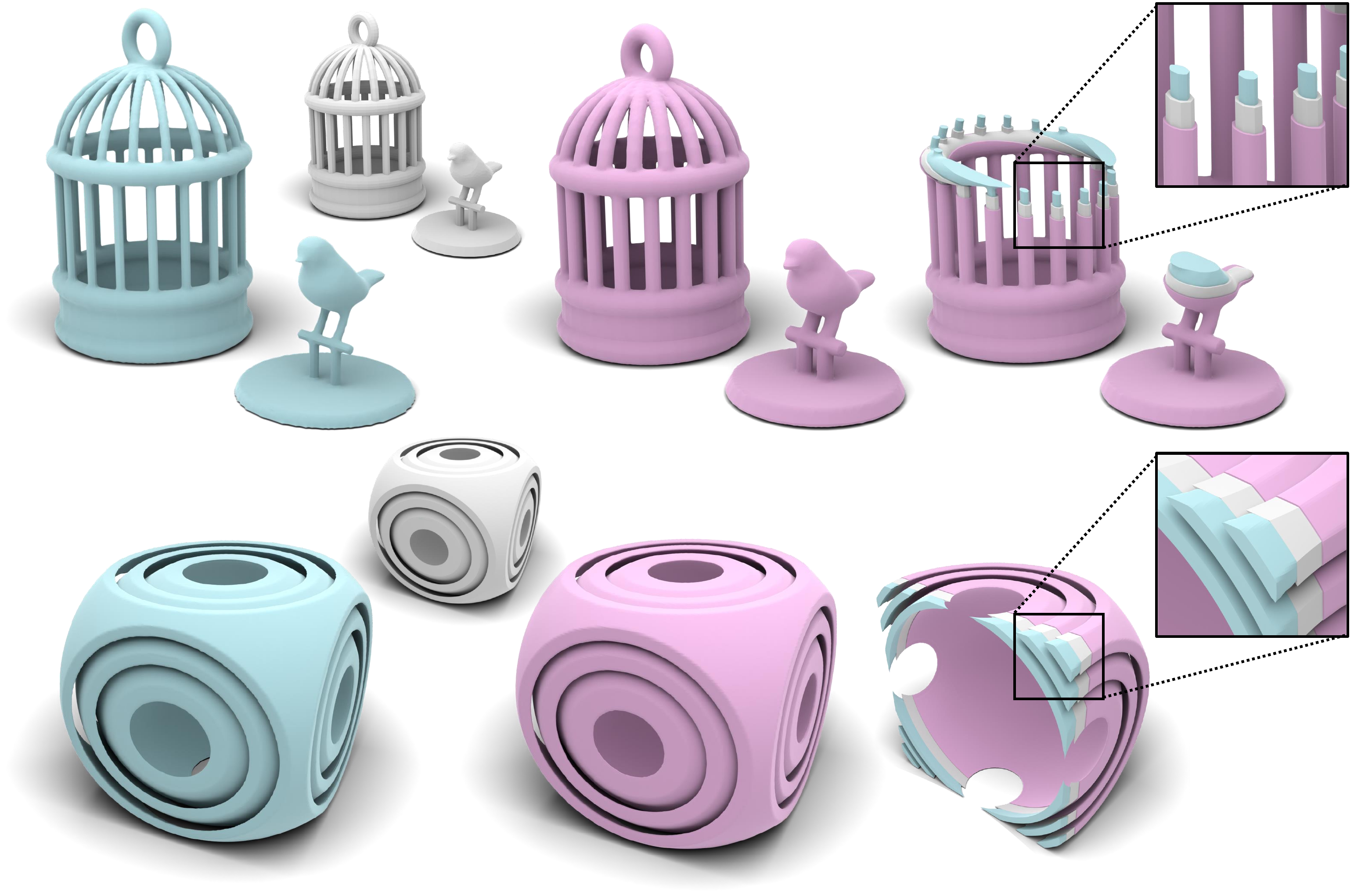}
	\vspace{-0.1in} \caption{
		ITS for models with multiple disconnected components. Top: the model with two distinct components. Bottom: the model with three distinct components. The original inputs are colored in grey. From left to right: the inner bounding surface, the outer surface, and the overlay visualization. 
		}
		\label{fig:disconnected}
\end{figure}

%% file: FigTex/FigGenus.tex
\begin{figure}
    \centering
    \includegraphics[width=8.5cm]{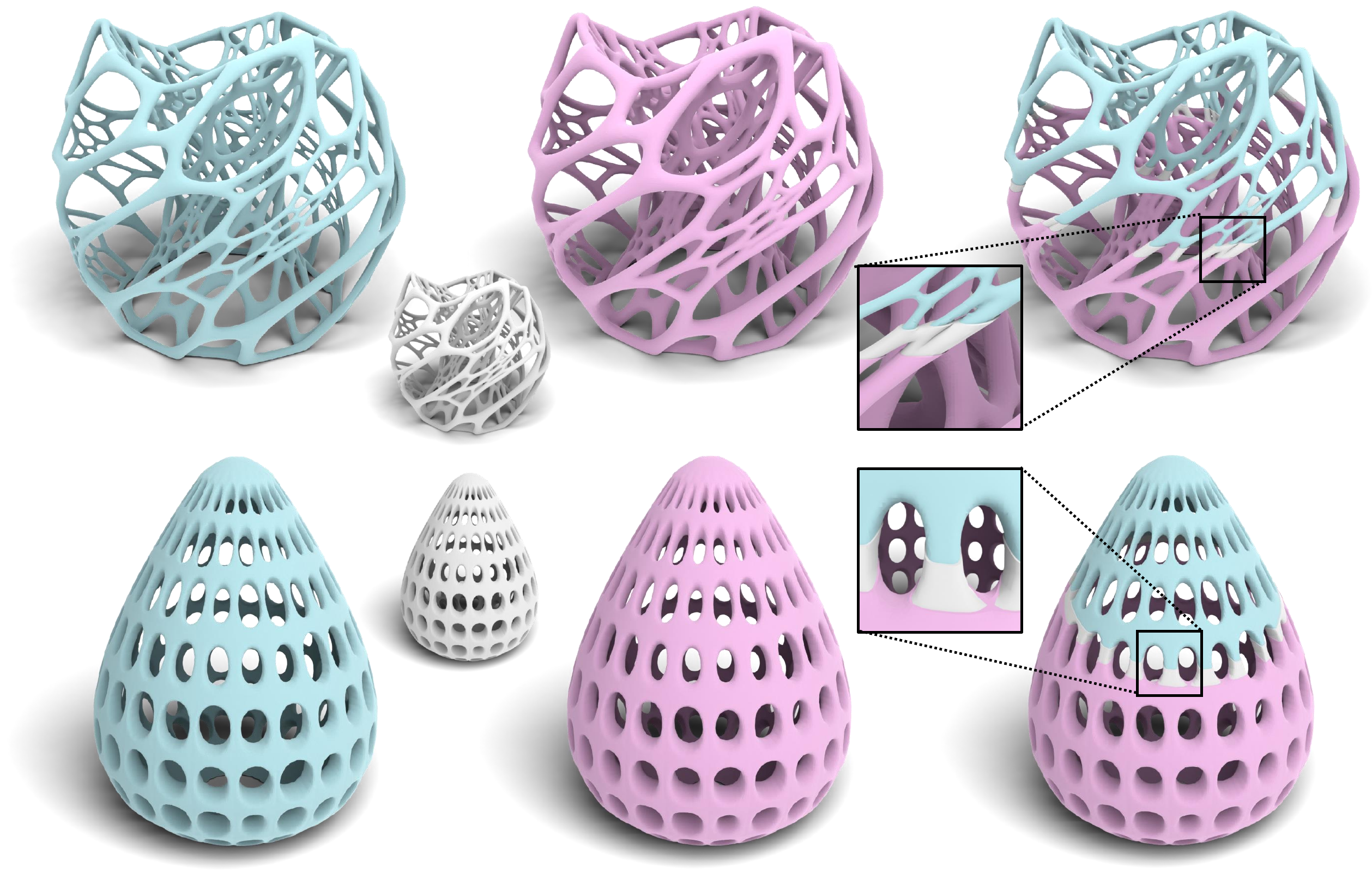}
    \vspace{-0.1in} \caption{
    ITS for models with high genus and thin structures.
    }
    \label{fig:genus}
\end{figure}

%% file: FigTex/FigDefect.tex
\begin{figure}
	\graphicspath{{Figures/}}
	\centering
	\begin{overpic}
		[width=8.5cm]{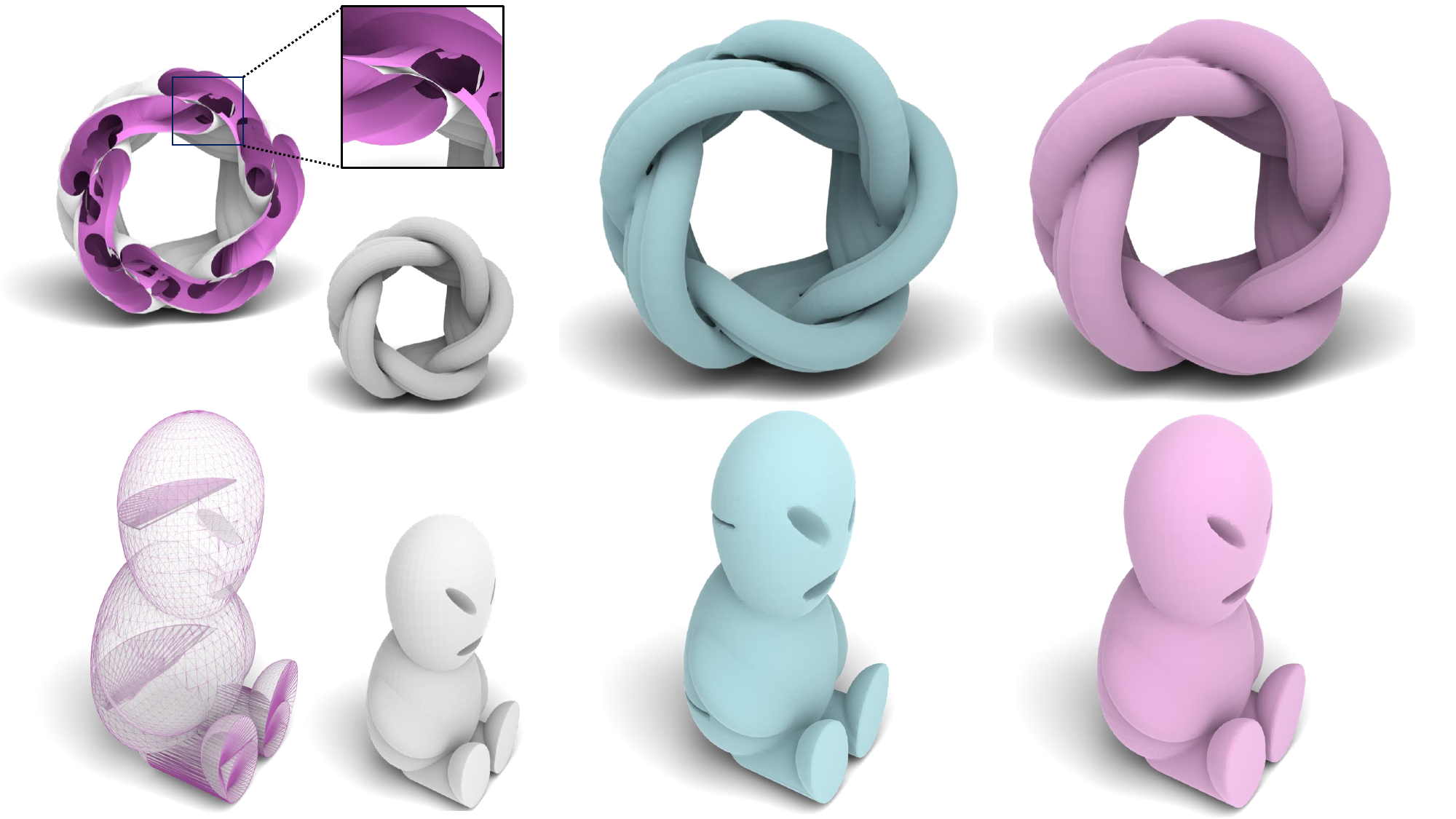}
	\end{overpic}
	\caption{
		Stability under defect-laden inputs. Top: a model with isolated facets and self-intersecting elements; bottom: a model with duplicated facets and non-manifold edges. 
		The bounding surfaces are generated using our method with~$\delta = 7$.
	}
	\label{fig:defect}
\end{figure}

%% file: FigTex/FigSoup.tex
\begin{figure}[!t]
\centering
    \graphicspath{{Figures/}}
    \includegraphics[width=8.5cm]{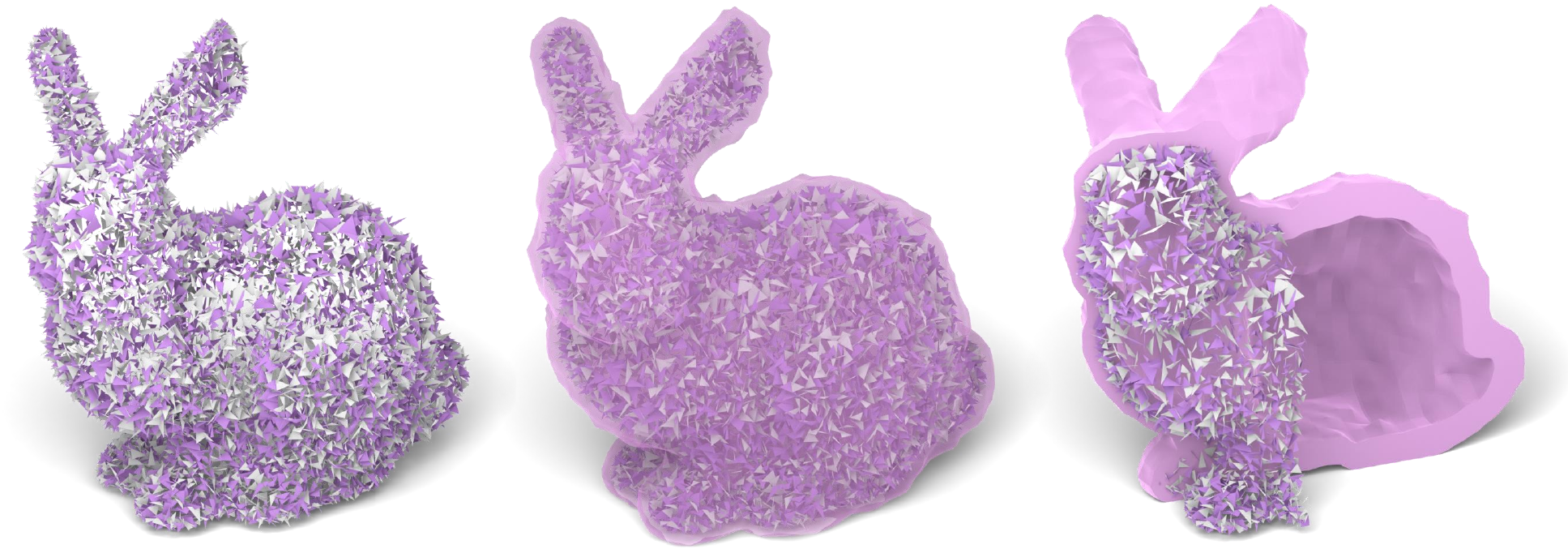}
    \caption{
    ITS for triangle soups, where $f$ is used to fit the UDF, instead of SDF. Left: A triangle soup. 
    Middle: Our outer shell can wrap the triangle soup.
    Right: Visualization of the double layers of the thin shell.
    Note that our thin shell can wrap the triangle soup rigorously.
    }
    \label{fig:soup}
\end{figure}

%% file: FigTex/FigCompare.tex
\begin{figure}
	\centering
	\begin{overpic}
		[width=8.5cm]{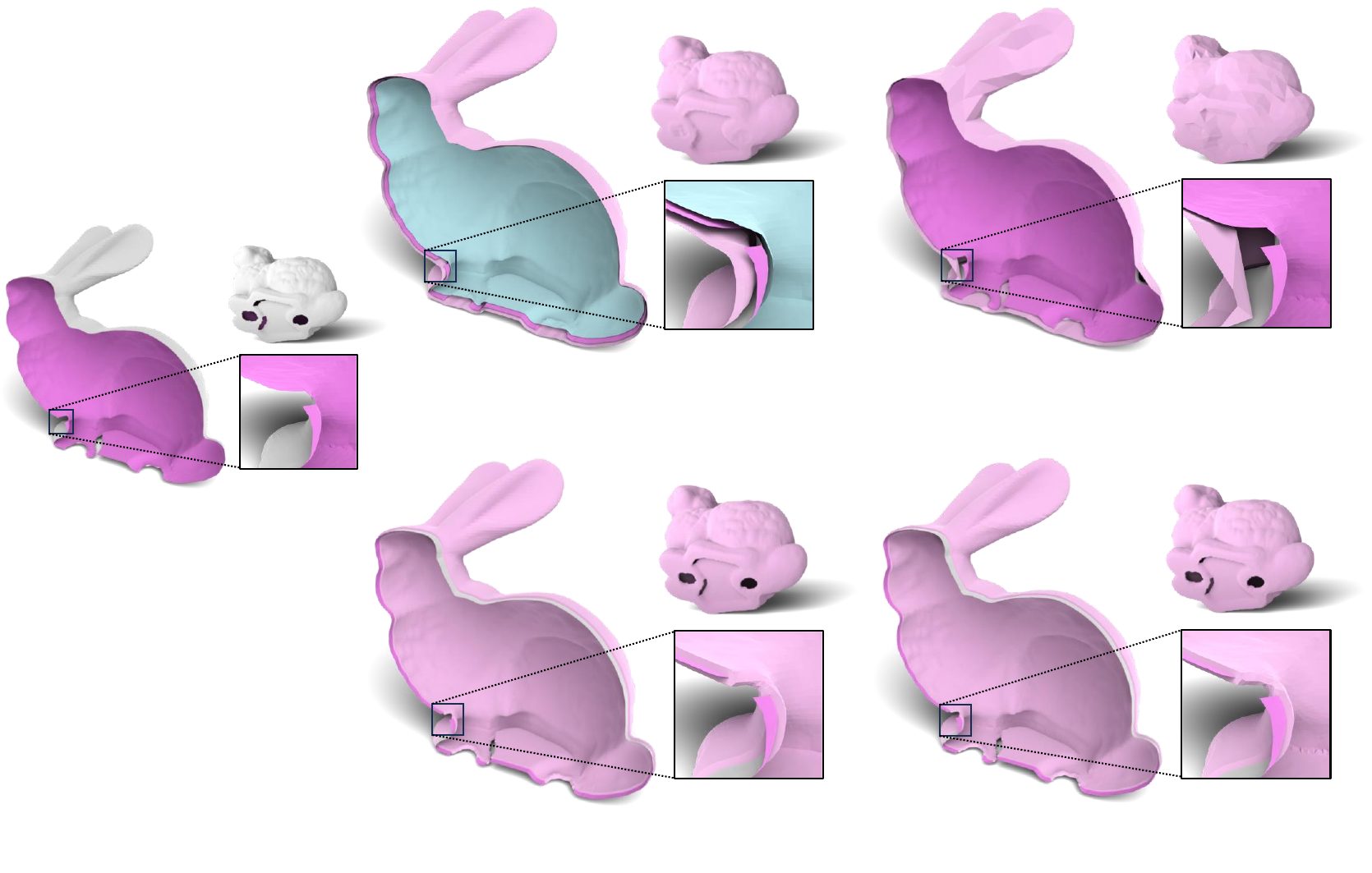}
		\put(12,24){Input}
		\put(40,34){Ours / SDF}
		\put(40,1){Ours / UDF}
		\put(76,34){AW~($\alpha = 1/30$)}
		\put(76,1){AW~($\alpha = 1/256$)}
	\end{overpic}
	\caption{In contrast to alpha wrapping~(AW), which can only generate an outer bounding surface, our algorithm is capable of generating both layers and filling open holes. 
	We configured~$K$ as~$8$ for our method and set the offset value to~$1/256$ for AW. To address open holes and retain intricate details from the input, we applied alpha values of~$1/30$ (top right) and~$1/256$ (bottom right) for AW, respectively.}		
	\label{fig:compare}
\end{figure}
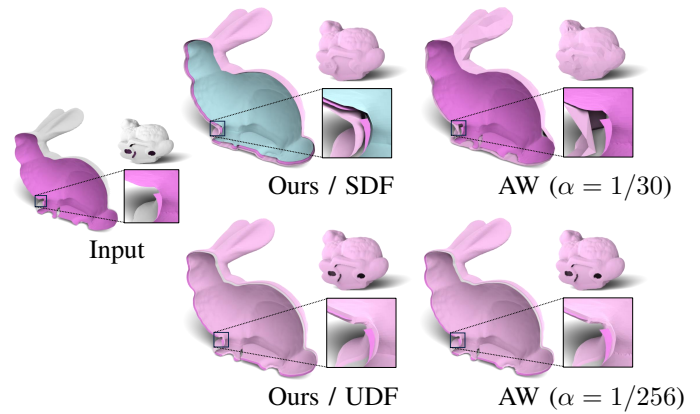

%% file: FigTex/FigQuery.tex
\begin{figure}
	\graphicspath{{Figures/Application/}}
	\centering
	\begin{overpic}
		[width=8.3cm]{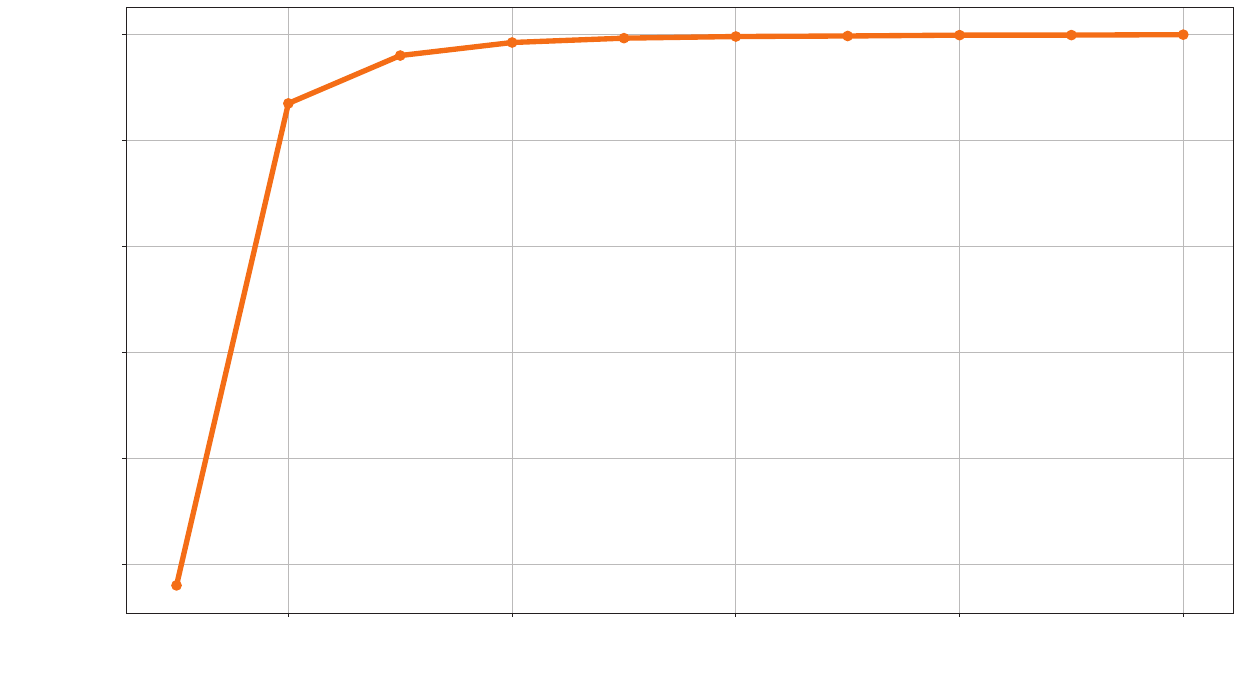}
        \put(-3, 4){\begin{turn}{90}Accuracy Ratio~(Ours / FCPW) \end{turn}}
        
		\put(1.7, 52.0){100}
		\put(2.0, 43.6){99.8}
		\put(2.0, 35.2){99.6}
		\put(2.0, 26.6){99.4}
		\put(2.0, 18.1){99.2}
		\put(2.0, 9.6){99.0}
  
		\put(22.3, 3){2}
		\put(40.3, 3){4}
		\put(59.0, 3){6}
		\put(76.8, 3){8}
		\put(93.3, 3){10}

		\put(40, -1){Size of Sampling Box}
	\end{overpic}
	\caption{
		We use FCPW to report accurate inside-outside results. For our method, we randomly sampled~100K points within a box of varying sizes while maintaining a value of $K=6$. Our results show that, for most practical situations where the query points are located within a bounding box with dimensions ranging from 1$\times$ to 10$\times$, our method can provide reliable query results. However, when query points are very close to the mesh surface, our method may become less accurate.
	}
	\label{fig:inside-outside-test}
\end{figure}

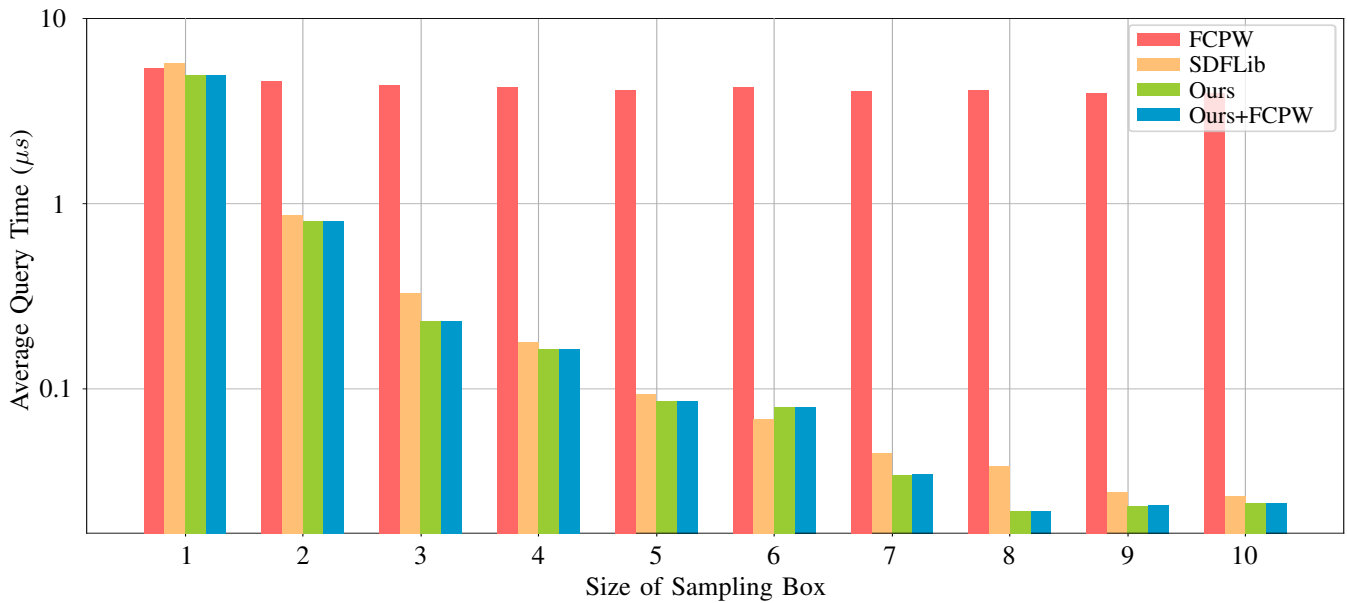
\begin{figure*}
	\graphicspath{{Figures/}}
	\centering
	\begin{overpic}
		[height=7cm, width=17cm]{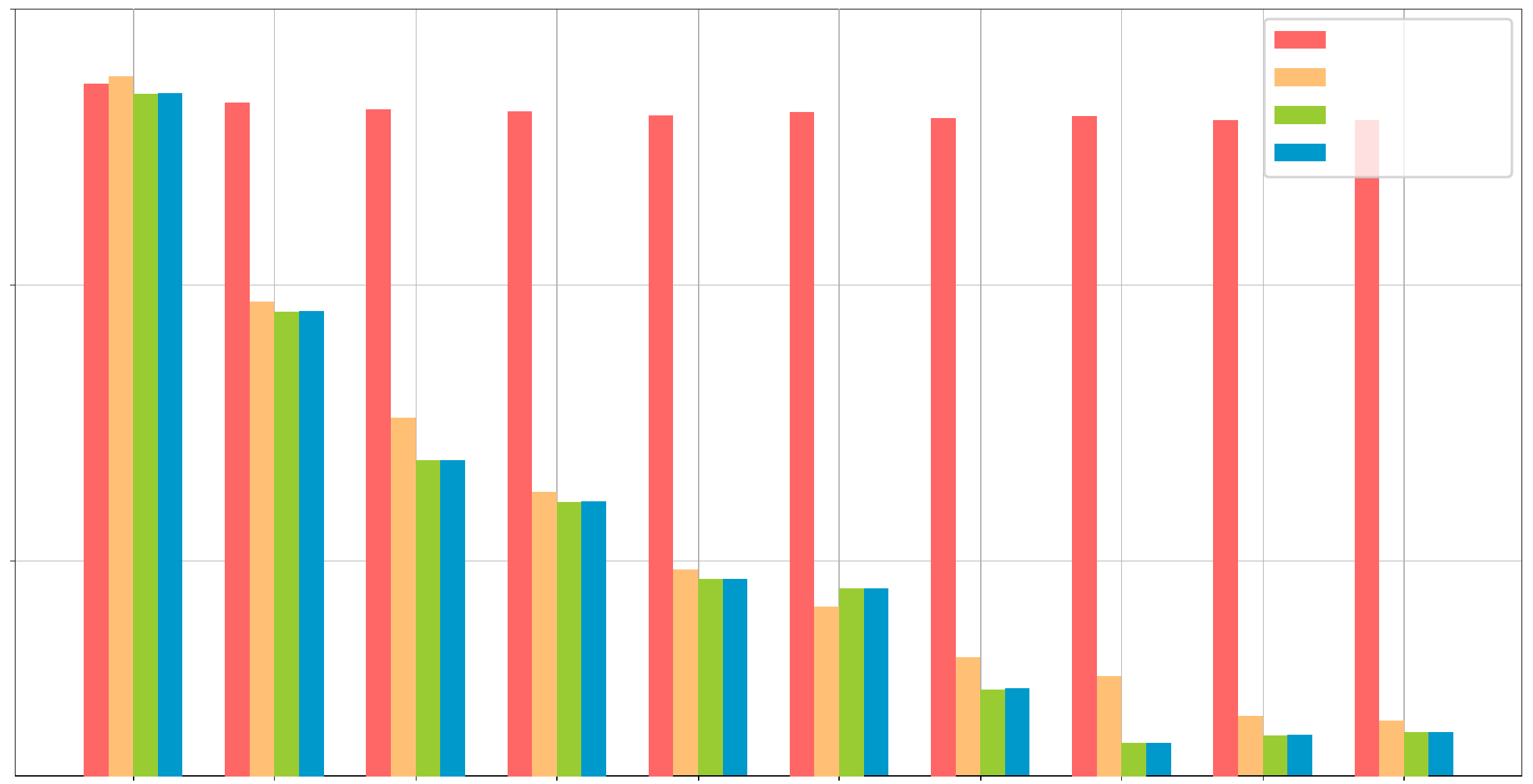}
		\put(-5, 10){\begin{turn}{90}Average Query Time~($\mu s$) \end{turn}}

        \put(87.2, 38.5){\small{FCPW}}
        \put(87.2, 36.5){\small{SDFLib}}
		\put(87.2, 34.5){\small{Ours}}
		\put(87.2, 32.5){\small{Ours+FCPW}}
        
		\put(-2.7, 40.0){10}
		\put(-1.7, 25.5){1}
		\put(-2.7, 11.1){0.1}

        \put(8.3, -2){1}
		\put(17.4, -2){2}
		\put(26.6, -2){3}
		\put(35.8, -2){4}
		\put(45.0, -2){5}
		\put(54.2, -2){6}
		\put(63.3, -2){7}
		\put(72.6, -2){8}
		\put(81.9, -2){9}
		\put(90.5, -2){10}

		\put(40, -4.4){Size of Sampling Box}
	\end{overpic}
	\vspace{0.6cm}
	\caption{
		Our approach produces a faithful inside-outside query result if the function value is not in the range of~$[\epsilon_1,\epsilon_2]$. Only when the function value is between~$\epsilon_1$ and $\epsilon_2$, we either return "OnSurface" or call FCPW to help report the true query result. The former strategy is numerically approximate, but the latter one is accurate (given the high accuracy of FCPW). We set~$K=6$ for our method and the SIMD width to 4 in FCPW. Tests on the Sapphos-Head model confirm the effectiveness of both strategies, demonstrating a substantial increase in speed compared to FCPW. 
	}
	\label{fig:inside-outside-time}
\end{figure*}

%% file: FigTex/FigSimplification.tex
\begin{figure}
    \graphicspath{{Figures/simplification/}}
 \begin{overpic}
		[width=8.5cm]{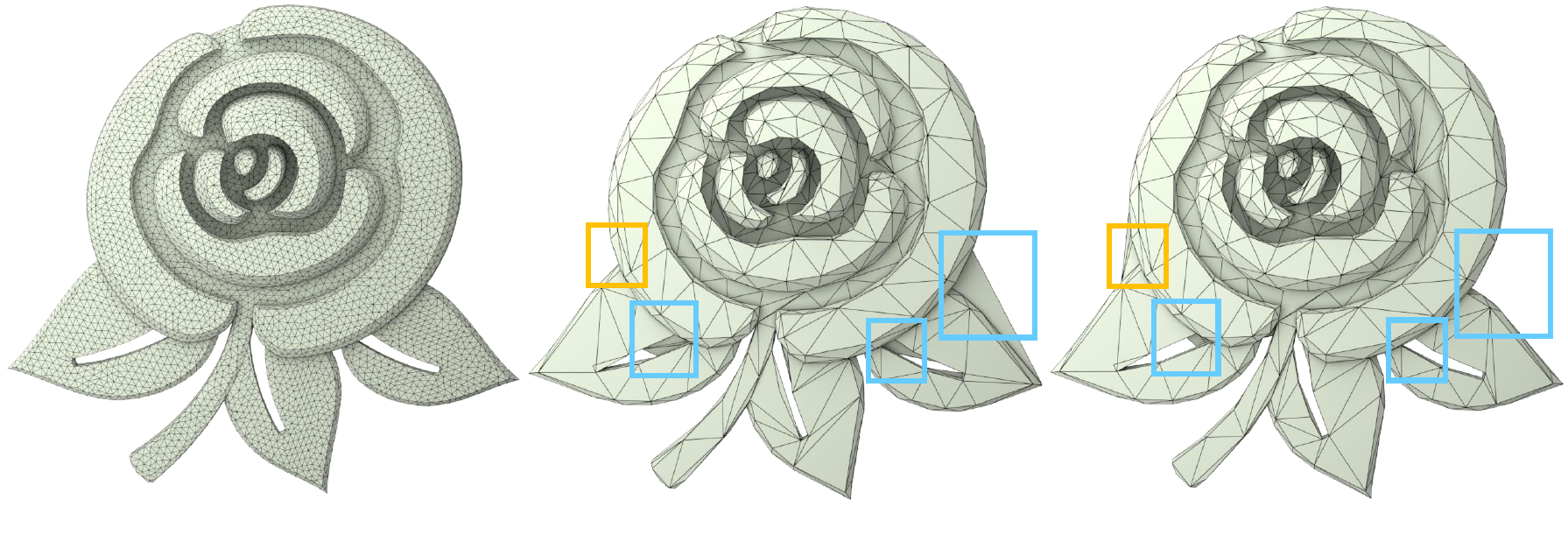}
		\put(9.7, -0.5){Input}
		\put(45.9, -0.5){QEM}
		\put(80.0, -0.5){Ours}
	\end{overpic}
	\caption{
  Apply the thin shell constraints with~$K=5$ to QEM  for an input mesh consisting of~17120 faces. When an optimal contraction is outside the thin-shells, we skip the edge collapse operation. This approach yields a mesh with fewer errors than QEM, as highlighted within the square frames. The simplified meshes both have 1209 faces.
  }
	\label{fig:simPara}
\end{figure}

\begin{figure}
    \graphicspath{{Figures/simplification/}}
    \centering
    \begin{overpic}
		[width=8.5cm]{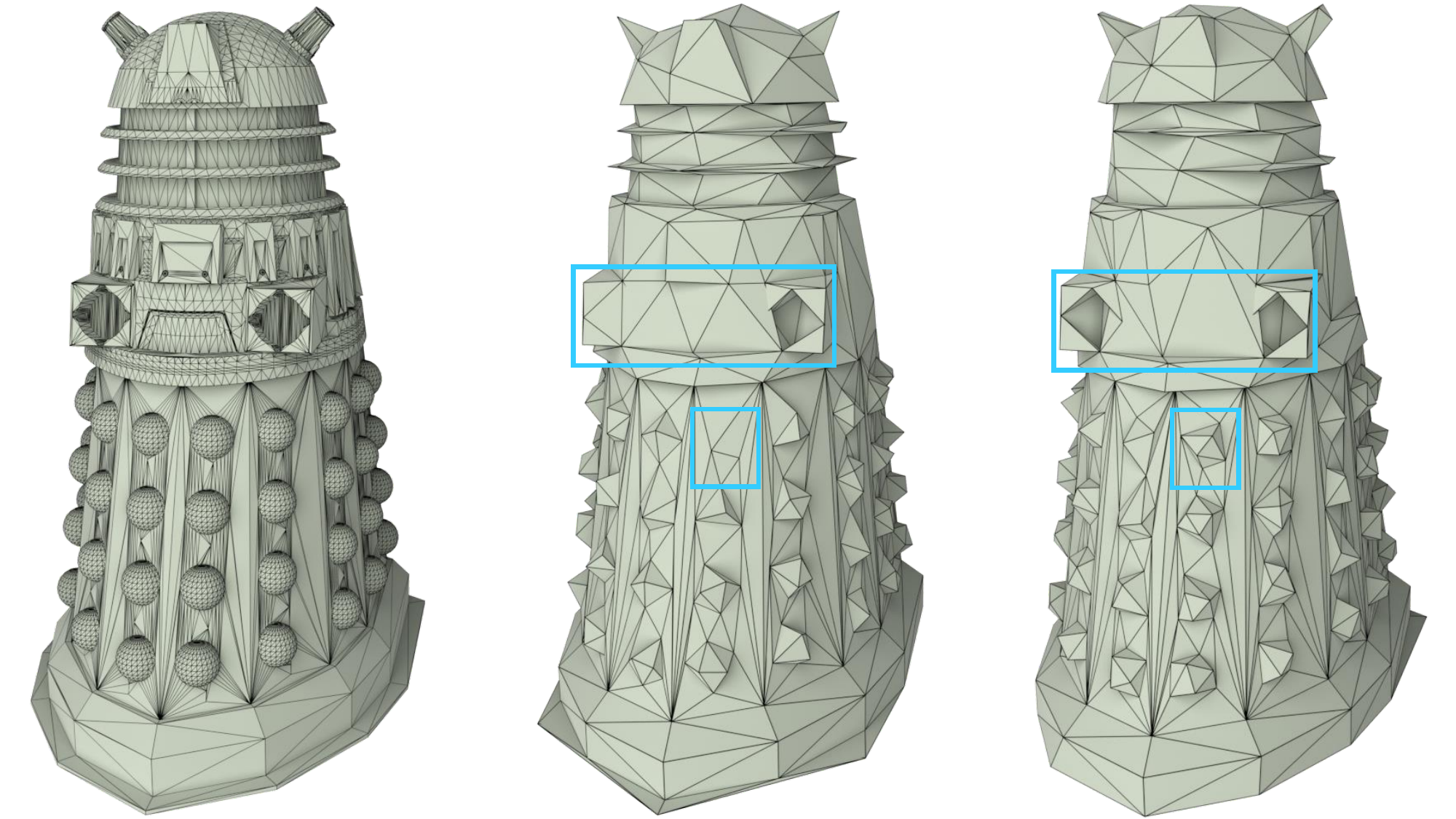}
		\put(8.7, -4){Input}
		\put(45.3, -4){QEM}
		\put(77.7, -4){Ours}
	\end{overpic}
\vspace{0.1in}\caption{
Comparison with QEM for reducing the triangle count from 30580 to 1223 using the parameter~$K=6$ of ITS. The introduction of the global term modifies the heap order, leading to a more simplified mesh approximation of the input. In contrast, QEM falls short in representing some significant features. Note that they have the same number of faces.
}
\label{fig:sim}
\end{figure}

%% file: main.bbl
\begin{thebibliography}{10}
\providecommand{\url}[1]{#1}
\csname url@samestyle\endcsname
\providecommand{\newblock}{\relax}
\providecommand{\bibinfo}[2]{#2}
\providecommand{\BIBentrySTDinterwordspacing}{\spaceskip=0pt\relax}
\providecommand{\BIBentryALTinterwordstretchfactor}{4}
\providecommand{\BIBentryALTinterwordspacing}{\spaceskip=\fontdimen2\font plus
\BIBentryALTinterwordstretchfactor\fontdimen3\font minus \fontdimen4\font\relax}
\providecommand{\BIBforeignlanguage}[2]{{%
\expandafter\ifx\csname l@#1\endcsname\relax
\typeout{** WARNING: IEEEtran.bst: No hyphenation pattern has been}%
\typeout{** loaded for the language `#1'. Using the pattern for}%
\typeout{** the default language instead.}%
\else
\language=\csname l@#1\endcsname
\fi
#2}}
\providecommand{\BIBdecl}{\relax}
\BIBdecl

\bibitem{97GarlandSimplification}
M.~Garland and P.~S. Heckbert, ``Surface simplification using quadric error metrics,'' in \emph{Proceedings of the 24th Annual Conference on Computer Graphics and Interactive Techniques}, ser. SIGGRAPH '97.\hskip 1em plus 0.5em minus 0.4em\relax USA: ACM Press/Addison-Wesley Publishing Co., 1997, pp. 209--216.

\bibitem{15MandadSimplification}
M.~Mandad, D.~Cohen-Steiner, and P.~Alliez, ``Isotopic approximation within a tolerance volume,'' \emph{ACM Transactions on Graphics (TOG)}, vol.~34, no.~4, pp. 1--12, 2015.

\bibitem{19polygonalShell}
F.~J. Melero, \'{A}ngel Aguilera, and F.~R. Feito, ``Fast collision detection between high resolution polygonal models,'' \emph{Computers \& Graphics}, vol.~83, pp. 97--106, 2019.

\bibitem{17boundaryProxy}
S.~Calderon and T.~Boubekeur, ``Bounding proxies for shape approximation,'' \emph{ACM Transactions on Graphics (TOG)}, vol.~36, no.~4, July 2017.

\bibitem{23Shell}
\BIBentryALTinterwordspacing
D.~Zint, N.~Maruani, M.~Rouxel-Labb{\'e}, and P.~Alliez, ``Feature-preserving offset mesh generation from topology-adapted octrees,'' \emph{Computer Graphics Forum}, vol.~42, no.~5, p.~12, 2023. [Online]. Available: \url{https://inria.hal.science/hal-04135266}
\BIBentrySTDinterwordspacing

\bibitem{15Cage}
L.~Sacht, E.~Vouga, and A.~Jacobson, ``Nested cages,'' \emph{ACM Transactions on Graphics (TOG)}, vol.~34, no.~6, November 2015.

\bibitem{23curvedesign}
\BIBentryALTinterwordspacing
R.~Xu, Y.~Jin, H.~Zhang, Y.~Zhang, Y.-k. Lai, Z.~Zhu, and F.-L. Zhang, ``A variational approach for feature-aware b-spline curve design on surface meshes,'' \emph{The Visual Computer}, vol.~39, no.~8, pp. 3767--3781, Aug 2023. [Online]. Available: \url{https://doi.org/10.1007/s00371-023-03001-x}
\BIBentrySTDinterwordspacing

\bibitem{17Printing}
M.~Yao, Z.~Chen, W.~Xu, and H.~Wang, ``Modeling, evaluation and optimization of interlocking shell pieces,'' \emph{Computer Graphics Forum}, vol.~36, no.~7, pp. 1--13, 2017.

\bibitem{07YeShellGeneration}
K.~Ye, K.~Zhou, Z.~Pan, Y.~Tong, and B.~Guo, ``Low distortion shell map generation,'' in \emph{2007 IEEE Virtual Reality Conference}, 2007, pp. 203--208.

\bibitem{23pattern-engraving}
J.~Hu, S.~Wang, Y.~He, Z.~Luo, N.~Lei, and L.~Liu, ``A parametric design method for engraving patterns on thin shells,'' \emph{IEEE Transactions on Visualization and Computer Graphics}, pp. 1--12, 2023.

\bibitem{15Mapping}
S.~Z. Kovalsky, N.~Aigerman, R.~Basri, and Y.~Lipman, ``Large-scale bounded distortion mappings,'' \emph{ACM Transactions on Graphics (TOG)}, vol.~34, no.~6, November 2015.

\bibitem{15narrowbandcvt}
Y.-S. Leung, X.~Wang, Y.~He, Y.-J. Liu, and C.~C.~L. Wang, ``A unified framework for isotropic meshing based on narrow-band euclidean distance transformation,'' \emph{Computational Visual Media}, vol.~1, pp. 239--251, 2015.

\bibitem{23Cloth}
\BIBentryALTinterwordspacing
Y.~Chen, T.~Xie, C.~Yuksel, D.~Kaufman, Y.~Yang, C.~Jiang, and M.~Li, ``Multi-layer thick shells,'' in \emph{ACM SIGGRAPH 2023 Conference Proceedings}, ser. SIGGRAPH '23.\hskip 1em plus 0.5em minus 0.4em\relax New York, NY, USA: Association for Computing Machinery, 2023. [Online]. Available: \url{https://doi.org/10.1145/3588432.3591489}
\BIBentrySTDinterwordspacing

\bibitem{04Shell}
J.~Peng, D.~Kristjansson, and D.~Zorin, ``Interactive modeling of topologically complex geometric detail,'' \emph{ACM Transactions on Graphics (TOG)}, vol.~23, no.~3, pp. 635--643, August 2004.

\bibitem{05ShellMap}
S.~D. Porumbescu, B.~Budge, L.~Feng, and K.~I. Joy, ``Shell maps,'' \emph{ACM Transactions on Graphics (TOG)}, vol.~24, no.~3, pp. 626--633, July 2005.

\bibitem{07HuangShell}
J.~Huang, X.~Liu, H.~Jiang, Q.~Wang, and H.~Bao, ``Gradient-based shell generation and deformation,'' \emph{Computer Animation and Virtual Worlds}, vol.~18, no. 4-5, pp. 301--309, 2007.

\bibitem{09Printing}
K.-Y. Lo, C.-W. Fu, and H.~Li, ``3{D} polyomino puzzle,'' \emph{ACM Transactions on Graphics (TOG)}, vol.~28, no.~5, pp. 1--8, December 2009.

\bibitem{19CurveDesign}
Y.~Jin, D.~Song, T.~Wang, J.~Huang, Y.~Song, and L.~He, ``A shell space constrained approach for curve design on surface meshes,'' \emph{Computer-Aided Design}, vol. 113, pp. 24--34, 2019.

\bibitem{20Shell}
Z.~Jiang, T.~Schneider, D.~Zorin, and D.~Panozzo, ``Bijective projection in a shell,'' \emph{ACM Transactions on Graphics (TOG)}, vol.~39, no.~6, November 2020.

\bibitem{22Alpha}
C.~Portaneri, M.~Rouxel-Labb\'{e}, M.~Hemmer, D.~Cohen-Steiner, and P.~Alliez, ``Alpha wrapping with an offset,'' \emph{ACM Transactions on Graphics (TOG)}, vol.~41, no.~4, July 2022.

\bibitem{02Shell}
C.~L. Bajaj, G.~Xu, R.~J. Holt, and A.~N. Netravali, ``Hierarchical multiresolution reconstruction of shell surfaces,'' \emph{Computer Aided Geometric Design}, vol.~19, no.~2, pp. 89--112, 2002.

\bibitem{04ImplicitShell}
C.~Shen, J.~F. O'Brien, and J.~R. Shewchuk, ``Interpolating and approximating implicit surfaces from polygon soup,'' in \emph{ACM SIGGRAPH 2004 Papers}, ser. SIGGRAPH '04.\hskip 1em plus 0.5em minus 0.4em\relax New York, NY, USA: Association for Computing Machinery, 2004, pp. 896--904.

\bibitem{11HanShell}
\BIBentryALTinterwordspacing
S.~Han, J.~Xia, and Y.~He, ``Constructing hexahedral shell meshes via volumetric polycube maps,'' \emph{Computer-Aided Design}, vol.~43, no.~10, pp. 1222--1233, 2011, solid and Physical Modeling 2010. [Online]. Available: \url{https://www.sciencedirect.com/science/article/pii/S0010448511001655}
\BIBentrySTDinterwordspacing

\bibitem{23MathMorphology}
\BIBentryALTinterwordspacing
V.~K. Suriyababu, C.~Vuik, and M.~M\"{o}ller, ``Towards a high quality shrink wrap mesh generation algorithm using mathematical morphology,'' \emph{Computer-Aided Design}, vol. 164, p. 103608, 2023. [Online]. Available: \url{https://www.sciencedirect.com/science/article/pii/S0010448523001409}
\BIBentrySTDinterwordspacing

\bibitem{20WangContainmentCheck}
\BIBentryALTinterwordspacing
B.~Wang, T.~Schneider, Y.~Hu, M.~Attene, and D.~Panozzo, ``Exact and efficient polyhedral envelope containment check,'' \emph{ACM Transactions on Graphics (TOG)}, vol.~39, no.~4, August 2020. [Online]. Available: \url{https://doi.org/10.1145/3386569.3392426}
\BIBentrySTDinterwordspacing

\bibitem{15_3L}
M.~Rouhani, A.~D. Sappa, and E.~Boyer, ``Implicit {B}-spline surface reconstruction,'' \emph{IEEE Transactions on Image Processing}, vol.~24, no.~1, pp. 22--32, 2015.

\bibitem{11SVO}
S.~Laine and T.~Karras, ``Efficient sparse voxel octrees,'' \emph{IEEE Transactions on Visualization and Computer Graphics}, vol.~17, no.~8, pp. 1048--1059, 2011.

\bibitem{10voxelization}
\BIBentryALTinterwordspacing
M.~Schwarz and H.-P. Seidel, ``Fast parallel surface and solid voxelization on gpus,'' \emph{ACM Transactions on Graphics (TOG)}, vol.~29, no.~6, 2010. [Online]. Available: \url{https://doi.org/10.1145/1882261.1866201}
\BIBentrySTDinterwordspacing

\bibitem{06PSR}
M.~Kazhdan, M.~Bolitho, and H.~Hoppe, ``Poisson surface reconstruction,'' in \emph{Proceedings of the Fourth Eurographics Symposium on Geometry Processing}, ser. SGP '06.\hskip 1em plus 0.5em minus 0.4em\relax Goslar, DEU: Eurographics Association, 2006, pp. 61--70.

\bibitem{04MLS}
N.~Amenta and Y.~J. Kil, ``Defining point-set surfaces,'' \emph{ACM Transactions on Graphics (TOG)}, vol.~23, no.~3, pp. 264--270, August 2004.

\bibitem{13Poisson}
M.~Kazhdan and H.~Hoppe, ``Screened {P}oisson surface reconstruction,'' \emph{ACM Transactions on Graphics (TOG)}, vol.~32, no.~3, July 2013.

\bibitem{17IHB}
M.~Pan, W.~Tong, and F.~Chen, ``Phase-field guided surface reconstruction based on implicit hierarchical {B}-splines,'' \emph{Computer Aided Geometric Design}, vol. 52--53, pp. 154--169, 2017, geometric Modeling and Processing 2017.

\bibitem{19Bspline}
M.~Kazhdan and H.~Hoppe, ``An adaptive multi-grid solver for applications in computer graphics,'' \emph{Computer Graphics Forum}, vol.~38, no.~1, pp. 138--150, 2019.

\bibitem{01RBF}
J.~C. Carr, R.~K. Beatson, J.~B. Cherrie, T.~J. Mitchell, W.~R. Fright, B.~C. McCallum, and T.~R. Evans, ``Reconstruction and representation of 3{D} objects with radial basis functions,'' in \emph{Proceedings of the 28th Annual Conference on Computer Graphics and Interactive Techniques}, ser. SIGGRAPH '01.\hskip 1em plus 0.5em minus 0.4em\relax New York, NY, USA: Association for Computing Machinery, 2001, pp. 67--76.

\bibitem{06RBF}
M.~Samozino, M.~Alexa, P.~Alliez, and M.~Yvinec, ``Reconstruction with {V}oronoi centered radial basis functions,'' in \emph{Proceedings of the Fourth Eurographics Symposium on Geometry Processing}, ser. SGP '06.\hskip 1em plus 0.5em minus 0.4em\relax Goslar, DEU: Eurographics Association, 2006, pp. 51--60.

\bibitem{10RBF}
J.~S{\" u}{\ss}muth, Q.~Meyer, and G.~Greiner, ``Surface reconstruction based on hierarchical floating radial basis functions,'' \emph{Computer Graphics Forum}, vol.~29, no.~6, pp. 1854--1864, 2010.

\bibitem{08wavelets}
J.~Manson, G.~Petrova, and S.~Schaefer, ``Streaming surface reconstruction using wavelets,'' in \emph{Proceedings of the Symposium on Geometry Processing}, ser. SGP '08.\hskip 1em plus 0.5em minus 0.4em\relax Goslar, DEU: Eurographics Association, 2008, pp. 1411--1420.

\bibitem{17wavelets}
J.~Horacsek and U.~Alim, ``Compactly supported biorthogonal wavelet bases on the body centered cubic lattice,'' \emph{Computer Graphics Forum}, vol.~36, no.~3, pp. 35--45, 2017.

\bibitem{05Fourier}
M.~Kazhdan, ``Reconstruction of solid models from oriented point sets,'' ser. SGP '05.\hskip 1em plus 0.5em minus 0.4em\relax Goslar, DEU: Eurographics Association, 2005, p. 73–es.

\bibitem{18Reconstruct}
W.~Lu, Z.~Shi, J.~Sun, and B.~Wang, ``Surface reconstruction based on the modified {G}auss formula,'' \emph{ACM Transactions on Graphics (TOG)}, vol.~38, no.~1, December 2018.

\bibitem{20Reconstruction}
P.~Erler, P.~Guerrero, S.~Ohrhallinger, N.~J. Mitra, and M.~Wimmer, ``Points2surf learning implicit surfaces from point clouds,'' in \emph{Computer Vision – ECCV 2020: 16th European Conference, Glasgow, UK, August 23–28, 2020, Proceedings, Part V}.\hskip 1em plus 0.5em minus 0.4em\relax Berlin, Heidelberg: Springer-Verlag, 2020, pp. 108--124.

\bibitem{22VoxelReconstruction}
H.~Li, X.~Yang, H.~Zhai, Y.~Liu, H.~Bao, and G.~Zhang, ``Vox-surf: Voxel-based implicit surface representation,'' \emph{IEEE Transactions on Visualization and Computer Graphics}, pp. 1--12, 2022.

\bibitem{22ISROverview}
S.~Lin, D.~Xiao, Z.~Shi, and B.~Wang, ``Surface reconstruction from point clouds without normals by parametrizing the {G}auss formula,'' \emph{ACM Transactions on Graphics (TOG)}, vol.~42, no.~2, October 2022.

\bibitem{18Deformation}
J.-M. Thiery, P.~Memari, and T.~Boubekeur, ``Mean value coordinates for quad cages in 3{D},'' \emph{ACM Transactions on Graphics (TOG)}, vol.~37, no.~6, December 2018.

\bibitem{DBLP:journals/tog/JuSW05}
T.~Ju, S.~Schaefer, and J.~D. Warren, ``Mean value coordinates for closed triangular meshes,'' \emph{{ACM} Trans. Graph.}, vol.~24, no.~3, pp. 561--566, 2005.

\bibitem{08JuCage}
T.~Ju, Q.-Y. Zhou, M.~van~de Panne, D.~Cohen-Or, and U.~Neumann, ``Reusable skinning templates using cage-based deformations,'' \emph{ACM Transactions on Graphics (TOG)}, vol.~27, no.~5, December 2008.

\bibitem{13YangCage}
X.~Yang, J.~Chang, R.~Southern, and J.~J. Zhang, ``Automatic cage construction for retargeted muscle fitting,'' \emph{The Visual Computer}, vol.~29, no.~5, pp. 369--380, May 2013.

\bibitem{09XianCage}
C.~Xian, H.~Lin, and S.~Gao, ``Automatic generation of coarse bounding cages from dense meshes,'' \emph{2009 IEEE International Conference on Shape Modeling and Applications}, pp. 21--27, 2009.

\bibitem{11XianCage}
H.~L. Chuhua~Xian and S.~Gao, ``Automatic cage generation by improved {OBB}s for mesh deformation,'' \emph{The Visual Computer}, vol.~28, pp. 21--33, 2011.

\bibitem{13StuartCage}
D.~A. Stuart, J.~A. Levine, B.~Jones, and A.~W. Bargteil, ``Automatic construction of coarse, high-quality tetrahedralizations that enclose and approximate surfaces for animation,'' in \emph{Proceedings of Motion on Games}, ser. MIG '13.\hskip 1em plus 0.5em minus 0.4em\relax New York, NY, USA: Association for Computing Machinery, 2013, pp. 213--222.

\bibitem{09BenChenCage}
M.~Ben-Chen, O.~Weber, and C.~Gotsman, ``Spatial deformation transfer,'' ser. SCA '09.\hskip 1em plus 0.5em minus 0.4em\relax New York, NY, USA: Association for Computing Machinery, 2009, pp. 67--74.

\bibitem{11DengCage}
Z.-J. Deng, X.-N. Luo, and X.-P. Miao, ``Automatic cage building with quadric error metrics,'' \emph{Journal of Computer Science and Technology}, vol.~26, no.~3, pp. 538--547, May 2011.

\bibitem{18LibiglWinding}
G.~Barill, N.~G. Dickson, R.~Schmidt, D.~I.~W. Levin, and A.~Jacobson, ``Fast winding numbers for soups and clouds,'' \emph{ACM Transactions on Graphics (TOG)}, vol.~37, no.~4, July 2018.

\bibitem{Eigen}
G.~Guennebaud, B.~Jacob \emph{et~al.}, ``Eigen: A {C}++ template library for linear algebra: Matrices, vectors, numerical solvers, and related algorithms,'' 2010, http://eigen.tuxfamily.org.

\bibitem{libigl}
A.~Jacobson, D.~Panozzo \emph{et~al.}, ``Libigl: A simple {C}++ geometry processing library,'' 2018, https://libigl.github.io/.

\bibitem{87mc}
\BIBentryALTinterwordspacing
W.~E. Lorensen and H.~E. Cline, ``Marching cubes: A high resolution 3{D} surface construction algorithm,'' \emph{ACM SIGGRAPH Computer Graphics}, vol.~21, pp. 163--169, August 1987. [Online]. Available: \url{https://doi.org/10.1145/37402.37422}
\BIBentrySTDinterwordspacing

\bibitem{Thingi10K}
Q.~Zhou and A.~Jacobson, ``Thingi10k: A dataset of 10,000 3{D}-printing models,'' \emph{arXiv preprint arXiv:1605.04797}, 2016.

\bibitem{15JiangBluenoise}
\BIBentryALTinterwordspacing
M.~Jiang, Y.~Zhou, R.~Wang, R.~Southern, and J.~J. Zhang, ``Blue noise sampling using an {SPH}-based method,'' \emph{ACM Transactions on Graphics}, vol.~34, no.~6, November 2015. [Online]. Available: \url{https://doi.org/10.1145/2816795.2818102}
\BIBentrySTDinterwordspacing

\bibitem{21FCPW}
R.~Sawhney, ``{FCPW}: Fastest closest points in the west,'' \url{https://github.com/rohan-sawhney/fcpw}, 2021.

\bibitem{23SDF}
\BIBentryALTinterwordspacing
E.~Pujol and A.~Chica, ``Triangle influence supersets for fast distance computation,'' \emph{Computer Graphics Forum}, vol.~42, no.~6, p. e14861, 2023. [Online]. Available: \url{https://onlinelibrary.wiley.com/doi/abs/10.1111/cgf.14861}
\BIBentrySTDinterwordspacing

\end{thebibliography}
